\documentclass[twocolumn]{aastex63}

\usepackage{amsmath}
\usepackage{amssymb}
\usepackage{graphicx}

\newcommand{\vecr}{\vec{R}}
\newcommand{\beq}{\begin{equation}}
\newcommand{\eeq}{\end{equation}}
\newcommand{\deriv}{\mathrm{d}}

\received{BLANK}
\revised{BLANK}
\accepted{BLANK}

\submitjournal{ApJ}
\shorttitle{Two-Point Separation Functions}
\shortauthors{Kervick, Walker,  Pe\~narrubia, Koposov}

\begin{document}

\title{Two-Point Separation Functions for Modeling Wide Binary Systems in Nearby Dwarf Galaxies}

\correspondingauthor{Christopher Kervick}
\email{ckervick@andrew.cmu.edu}

\author{Christopher Kervick}
\affiliation{McWilliams Center for Cosmology, Dept. of Physics, Carnegie Mellon University, 5000 Forbes Avenue, Pittsburgh, PA 15213, USA}

\author[0000-0003-2496-1925]{Matthew G. Walker}
\affiliation{McWilliams Center for Cosmology, Dept. of Physics, Carnegie Mellon University, 5000 Forbes Avenue, Pittsburgh, PA 15213, USA}

\author{Jorge Pe\~narrubia}
\affiliation{Institute for Astronomy, University of Edinburgh, Royal Observatory, Blackford Hill, Edinburgh EH9 3HJ, UK}
\affiliation{Center for Statistics, University of Edinburgh, School of Mathematics, Edinburgh, 
EH9 3FD, United Kingdom}

\author[0000-0003-2644-135X]{Sergey~E.~Koposov}
\affiliation{Institute for Astronomy, University of Edinburgh, Royal Observatory, Blackford Hill, Edinburgh EH9 3HJ, UK}
\affiliation{Institute of Astronomy, University of Cambridge, Madingley Road, Cambridge CB3 0HA, UK}

\begin{abstract}

We use a geometric method to derive (two-dimensional) separation functions amongst pairs of objects within populations of specified position function $\deriv N/\deriv\vecr$.  We present analytic solutions for separation functions corresponding to a uniform surface density within a circular field, a Plummer sphere (viewed in projection), and the mixture thereof---including contributions from binary objects within both sub-populations.  These results enable inferences about binary object populations via direct modeling of object position and pair separation data, without resorting to standard estimators of the  two-point correlation function.  Analyzing mock data sets designed to mimic known dwarf spheroidal galaxies, we demonstrate the ability to recover input properties including the number of wide binary star systems and, in cases where the number of resolved binary pairs is assumed to be $\ga$ a few hundred, characteristic features (e.g., steepening and/or truncation) of their separation function.  Combined with forthcoming observational capabilities, this methodology opens a window onto the formation and/or survival of wide binary populations in dwarf galaxies, and offers a novel probe of dark matter substructure on the smallest galactic scales. 
\end{abstract}


\section{Introduction} \label{sec:intro}

Wide binary star systems are vulnerable to disruption via encounters with perturbers \citep[e.g.][]{chandra44,heggie75,weinberg87,jiang10}, making them useful tracers of dark structure  \citep[e.g.][]{bahcall85,chaname04,yoo04}.  Usually detected as stellar pairs with common proper motion, Galactic wide binaries with separations $2\la \log_{10}(s/\mathrm{A.U.})\la 4$ are typically characterized using a power-law separation function, $p(s)\propto s^{-\gamma}$, with $\gamma \approx 1.0 - 1.6$ \citep{chaname04,lepine07,andrews17,el-badry18}.  At larger separations, the power-law slope appears to steepen by an amount that depends on age and kinematics \citep{tian2019separation}.  Characterization of the binary separation function in different environments is crucial for understanding the formation and destruction of these weakly bound systems \citep{moeckel10,kouwenhoven10,el-badry18,penarrubia20}.

Wide binaries within the Milky Way's nearest satellites can potentially trace low-mass dark matter halo structure and substructure.  The faintest dwarf galaxies have the largest mass-to-light ratios and dark matter densities known, with $M/L_V\sim 10^{2-4}$ in solar units and $\rho\ga 1 M_{\odot}$ pc$^{-3}$ within their halflight radii \citep{mateo98,mcconnachie12,simon19}.  In the scale-free hierarchy of structure expected under the cold dark matter paradigm, such large dark matter densities imply that of all galaxies, the faintest dwarfs host the highest number densities of subhalo (and sub-subhalo, etc.)  perturbers \citep[e.g.][]{springel08}. Indeed, N-body experiments demonstrate that, on ultrafaint dwarf galaxy scales, the wide binary separation function is strongly sensitive to both the structure of, and amount of substructure within, the host dark matter halo \citep{penarrubia10c,penarrubia2016wide}.      

It is already well known that the Milky Way's faint satellites contain significant numbers of \textit{short-}period binary stars ($P\la 100$ years).  Several multi-epoch spectroscopic studies have identified likely binary systems as radial velocity variables with amplitudes up to $\sim 10$ km s$^{-1}$, implying binary fractions of $f_{\rm b}\approx 0.5$ \citep{martinez11,koposov11,minor13,koch14,spencer18}.  Similar fractions are derived from deep \textit{Hubble Space Telescope} imaging of several ultrafaint dwarf galaxies, based on the tendency of unresolved pairs to broaden the main sequence redward and brighter \citep{geha13}.  However, these techniques necessarily miss wide binaries, which can have orbital periods exceeding thousands of years and may be resolved as point-source pairs.  Furthermore, finding wide binaries via common proper motions at dwarf galaxy distances ($\sim 20-100$ kpc) would require precision $\la 1$ milli-arcsec century$^{-1}$ at $\sim 20$th magnitude, two orders of magnitude finer than what is delivered by \textit{Gaia}'s EDR3 catalog \citep{gaia21}.   Thus empirical constraints on wide binary populations within dwarf galaxies are at present nonexistent.  

As a result, we do not know how or even whether wide binaries ever form within dwarf galaxies, much less how or whether they survive.  Possible formation mechanisms include the same ones invoked to explain wide binaries observed within the Galaxy, e.g. 1) gravitational entrapment of neighbors as star-forming regions expand in response either to rapid gas loss \citep{kouwenhoven10,moeckel10} or to collisional relaxation  \citep{moeckel11}, 2) three-body interactions in which one star is scattered to a large orbit but remains weakly bound to a relatively compact pair \citep{reipurth12}, 3) binding of adjacent pre-stellar cores that  move at slow relative  velocity within the star-forming cloud \citep{tokovinin17}, and/or 4) entrapment within the long-lived tails that emanate from tidally disrupted star clusters \citep{penarrubia20}.    

Of course, the only way to prove that wide binaries both form and survive within dwarf galaxies would be to find them.  Here we investigate the detectability of wide binaries in dwarf galaxies via single-epoch images that might resolve binary pairs.  For reference, at distance 100 kpc, a pair of objects separated by 0.1 pc subtends 0.2 arcsec---approximately four times the diffraction limit of the \textit{Hubble Space Telescope} and the upcoming \textit{Nancy Grace Roman Space Telescope}.  

The workhorse statistic for analyzing spatial inhomogeneities is the two-point correlation function (2PCF; e.g., \citealt{peebles80}), which characterizes object clustering as a function of spatial scale.  Specifically, the 2PCF quantifies the `excess' number of object pairs at a given separation with respect to the expectation for a random field.  The 2PCF is widely used, e.g., to infer cosmological parameters from observations of large-scale structure \citep[e.g.,][]{des21}.  More relevant to the present study, \citet{longhitano10} model the 2PCF of late-type stars in the solar neighborhood, reporting that $\sim 10\%$ of such stars belong to wide binary systems with projected separation $10^{-3}\la s/(\rm pc)\la 1$.

Standard estimators of the 2PCF from cosmological large-scale structure quantify the expected  `background' by averaging over large numbers of Monte Carlo realizations of the random field \citep[e.g.,][]{davis83,hamilton93,landy93}.  This procedure can faithfully account for features in the 2PCF that arise due to complications like survey footprint, incompleteness and known selection effects.  

For the purpose of studying wide binary stars in dwarf galaxies, here we take a different approach that exploits the relative simplicity of these systems.  For nearby dwarf galaxies, the random field of resolved stars is  generally well characterized by an analytic surface number density function that includes an approximately uniform foreground (Milky Way) component \citep[e.g.,][]{irwin95,martin08,moskowitz2020stellar}.  Here we develop formalism for calculating two-point separation functions directly from the surface number density function---including the contribution from binary objects.  We derive analytic results for the case in which the number density function is the mixture of a Plummer sphere and a uniform background---the scenario commonly invoked to model dwarf galaxy star counts.  While these results can provide random-field input to a standard 2PCF estimator, they also provide a means to infer properties of binary object populations via direct modeling of the empirical separation function.  We demonstrate the latter capability using mock imaging catalogs generated to mimic structural parameters observed for the Milky Way's known dwarf-galactic satellites, highlighting the conditions under which wide binary populations---if they exist---can reliably be detected and characterized within these systems.

\section{2D Separation Functions} \label{sec:method}

First we develop formalism for calculating 2D separation functions.  Consider a population of objects whose 2D (projected onto the plane of the sky) positions, $\vecr\in\mathbb{R}^2$, are distributed randomly, via Poisson point process, according to surface number density function $\Sigma(\vecr)\equiv \deriv N/\deriv\vecr$.  The total number of objects has expectation value $\langle N\rangle=\iint \Sigma(\vecr)\,\deriv\vecr$.  The probability density of object positions is then $p(\vecr)=\Sigma(\vecr)/\langle N\rangle$.  
We seek to calculate the  probability density for separations between pairs of objects drawn randomly from the population.

\begin{figure}
    \includegraphics[width=3.5in,trim=0.3in 0.3in 0.3in 1.3in, clip]{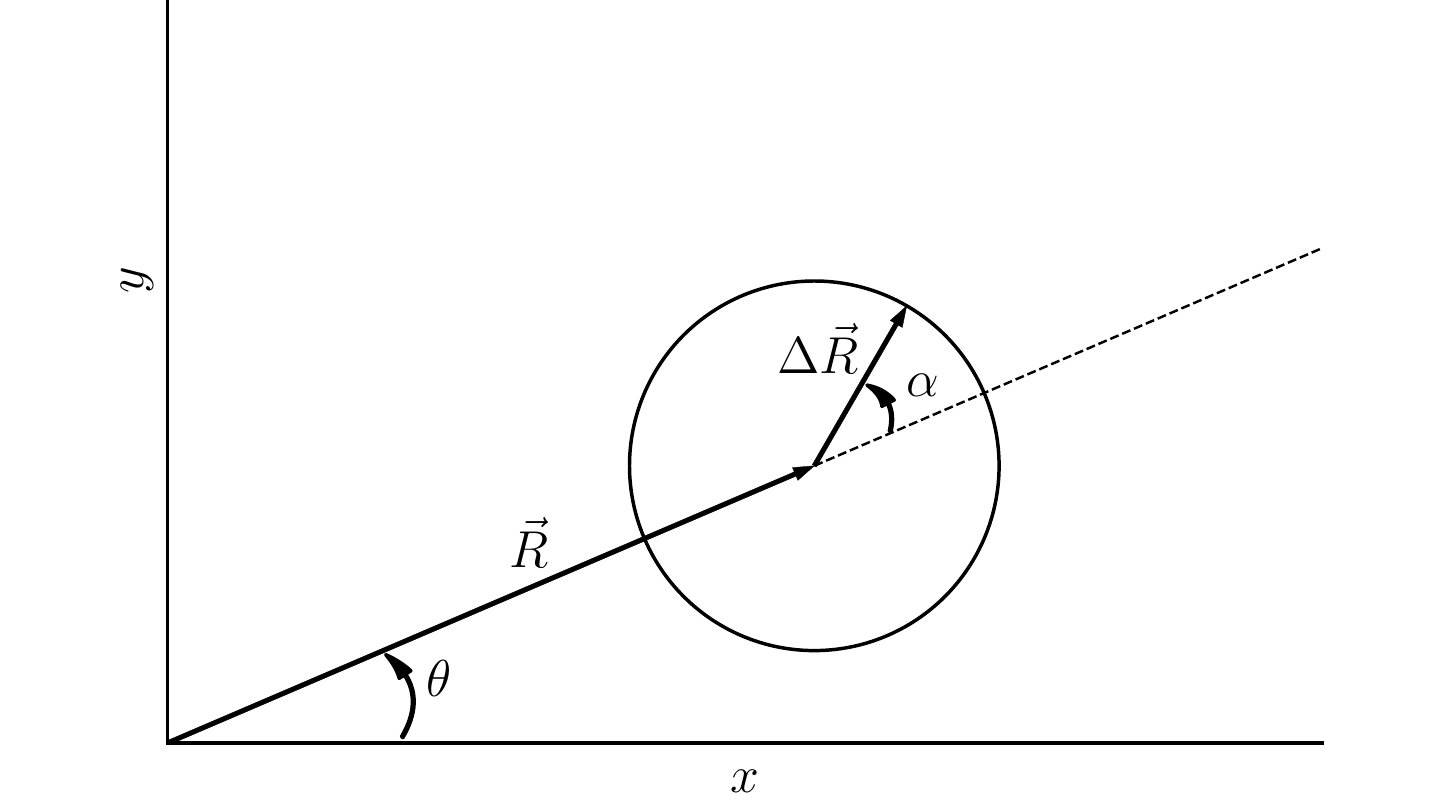}
    \caption{Geometry for calculating probability density for separations of objects from position $\vecr$ (see Equation \ref{eq:mu_general}).
\label{fig:wb_geometry}}
\end{figure}

Figure \ref{fig:wb_geometry} illustrates the relevant geometry.  Letting $\alpha$ be the angle between $\vecr$ and the vector displacement to a second position $\vecr+\Delta\vecr$, the number of objects that are separated from $\vecr$ by projected distance within the interval $s,s+\mathrm{d}s$ is $\mu(s|\vecr)\,\mathrm{d}s$, where
\beq \label{eq:mu_general}
    \mu(s|\vecr)\equiv \frac{\deriv N}{\deriv s}\biggr |_{\vecr}=\int_0^{2\pi}\Sigma(\vecr+\Delta\vecr)\,s\,\mathrm{d}\alpha,
\eeq
is the conditional separation function, $\Delta\vecr=s\bigl (\cos(\theta+\alpha)\hat{x}+\sin(\theta+\alpha)\hat{y}\bigr )$ and $\theta$ is the position angle at $\vecr$.  The number of pairs having one object within position interval $\vecr$, $\vecr+\deriv\vecr$ and separation within the interval $s$, $s+\deriv s$ is $\psi(s,\vecr)\,\deriv s\,\deriv \vecr$, where
\beq
    \psi(s,\vecr)\equiv \frac{\deriv N}{\deriv\vecr\,\deriv s}=\Sigma(\vecr)\,\mu(s|\vecr)
    \label{eq:psi}
\eeq
is the joint position-separation function.  
The number of pairs having separation within the interval $s,s+\deriv s$, regardless of position, is $\phi(s)\,\deriv s$, where 
\beq
    \phi(s)\equiv \frac{\deriv N}{\deriv s}=\iiint \psi(s,\vecr)\,\deriv\vecr
    \label{eq:phi}
\eeq
is the marginal separation function.  

For the purpose of calculating normalized probability densities, these separation functions integrate to $\int \mu(s|\vecr)\,\deriv s=\langle N\rangle$ and $\iiint \phi(s,\vecr)\,\deriv s\,\deriv\vecr=\int \phi(s)\,\deriv s=\langle N\rangle^2$. 

\subsection{Mixtures}
\label{subsec:mixtures}
It is useful to generalize to the case in which the population of objects comprises a mixture of $N_{\rm pop}$ distinct sub-populations that each, independently, follow distinct spatial distributions---i.e., $\Sigma_{\rm mix}(\vecr)=\sum_{i=1}^{N_{\rm pop}}\Sigma_i(\vecr)$.  The total number of objects has expectation value $\langle N_{\rm mix}\rangle=\iint \Sigma_{\rm mix}(\vecr)\,\deriv\vecr=\sum_{i=1}^{N_{\rm pop}}\langle N_i\rangle$, where $\langle N_i\rangle=\iint\Sigma_i(\vecr)\,\deriv\vecr$ is the expectation value for the number of objects in the $i^{\rm th}$ sub-population.  The conditional separation function is 
\beq
    \mu_{\rm mix}(s|\vecr)=\sum_{i=1}^{N_{\rm pop}}\mu_i(s|\vecr),
    \label{eq:mix_conditional}
\eeq
where $\mu_i(s|\vecr)=\int_{0}^{2\pi}\Sigma_i(\vecr+\Delta\vecr)\,s\,\deriv\alpha$.

The joint position-separation function is
\begin{equation}
    \psi_{\rm mix}(s,\vecr)=\sum_{i=1}^{N_{\rm pop}}\sum_{j=1}^{N_{\rm pop}} \psi_{i,j}(s,\vecr),
\end{equation}
where $\psi_{i,j}(s,\vecr)\equiv \Sigma_{i}(\vecr)\,\mu_j(s|\vecr)$.  
The marginal separation function is 
\beq
    \phi_{\rm mix}(s)=\sum_{i=1}^{N_{\rm pop}}\sum_{j=1}^{N_{\rm pop}}\phi_{i,j}(s),
    \label{eq:mix_marginal}
\eeq
where $\phi_{i,j}(s)\equiv \iint\psi_{i,j}(s,\vecr)\,\deriv\vecr$.  The cross terms are symmetric, such that $\phi_{i,j}(s)=\phi_{j,i}(s)$.

\subsection{Binaries} \label{subsec:binaries}

We now consider the separation functions that pertain to a population within which some fraction of the objects are split into binary systems that, at position $\vecr$, follow internal separation function $\phi_{\rm b}(s|\vecr)$, normalized so that $\int\phi_{\rm b}(s|\vecr)\,\deriv s=1$.  The surface number density can be written as the sum of contributions from single objects and binary objects: $\Sigma(\vecr)=\Sigma_{\rm s}(\vecr)+\Sigma_{\rm b}(\vecr)$.  We define the binary fraction as $f_{\rm b}(\vecr)\equiv \Sigma_{\rm b}(\vecr)/\Sigma(\vecr)$, such that while $\Sigma(\vecr)$ remains the surface number density of statistically-independent  objects, the surface number density of \textit{countable items} is $\Sigma'(\vecr)=\bigl (1+f_{\rm b}(\vecr)\bigr )\Sigma(\vecr)$---i.e., a binary system is one object that comprises two countable items.  The number of countable items has expectation value $\langle N'\rangle = \iint \Sigma'(\vecr)\,\deriv\vecr=(1+f_{\rm b})\langle N\rangle$, where $f_{\rm b}\equiv \langle N\rangle^{-1}\iint f_{\rm b}(\vecr)\,\Sigma(\vecr)\,\deriv\vecr$ is the globally-averaged binary fraction.

The joint position-separation function for countable items is (see derivation in Appendix \ref{app:1})
\beq
    \psi'(s,\vecr)\approx\bigl (1+f_{\rm b}(\vecr)\bigr )^2\psi(s,\vecr)+2\,f_{\rm b}(\vecr)\,\Sigma(\vecr)\,\phi_{\rm b}(s,\vecr),
    \label{eq:psiprime}
\eeq
where $\psi(s,\vecr)$ is the joint position-separation function for independent objects.

The conditional separation function for countable items is then
\begin{align}
    \mu'(s|\vecr)=\frac{\psi'(s,\vecr)}{\Sigma'(\vecr)}\hspace{2in}\nonumber\\
    \approx\bigl (1+f_{\rm b}(\vecr)\bigr )\,\mu(s|\vecr)+\frac{2f_{\rm b}(\vecr)}{1+f_{\rm b}(\vecr)}\,\phi_{\rm b}(s|\vecr)
    \label{muprime}
\end{align}
where $\mu(s|\vecr)$ is the conditional separation function for independent objects.  

The marginal separation function for countable items is $\phi'(s)=\iint\psi'(s,\vecr)\,\deriv\vecr$.  If the binary fraction and internal separation function are both independent of position, such that $f_{\rm b}(\vecr)\rightarrow f_{\rm b}$ and $\phi_{\rm b}(s|\vecr)\rightarrow \phi_{\rm b}(s)$, then the marginal separation function for countable items simplifies to
\beq
    \phi'(s)\approx (1+f_{\rm b})^2\,\phi(s)+2\,f_{\rm b}\langle N\rangle\,\phi_{\rm b}(s),
\eeq
where $\phi(s)$ is the marginal separation function for independent objects.

Also, if the binary fraction and internal separation function are independent of position, then the separation functions for countable items integrate to
\begin{align}
    \int\mu'(s|\vecr)\,\deriv s\approx (1+f_{\rm b})\langle N\rangle +\frac{2f_{\rm b}}{1+f_{\rm b}};\hspace{1in}\nonumber\\
    \iiint \psi'(s,\vecr)\,\deriv s\,\deriv\vecr=\int\phi'(s)\,\deriv s\hspace{1.5in}\nonumber\\
    \approx (1+f_{\rm b})^2\langle N\rangle^2+2\,f_{\rm b}\langle N\rangle.\hspace{0.75in}
\end{align}

\subsection{Mixtures and binaries}
In the most general case that we consider here, the population consists of a mixture of $N_{\rm pop}$ distinct sub-populations, each independently following its own spatial distribution and each containing binary systems that follow independent internal separation functions.  The total surface density is  $\Sigma_{\rm mix}(\vecr)=\sum_{i=1}^{N_{\rm pop}}\Sigma_i(\vecr)=\sum_{i=1}^{N_{\rm pop}}\Sigma_{{\rm s}_i}(\vecr)+\Sigma_{{\rm b}_i}(\vecr)$, where $\Sigma_{{\rm s}_i}(\vecr)$ and $\Sigma_{{\rm b}_i}(\vecr)$ are surface number densities of single and binary objects, respectively, within the $i^{\rm th}$ sub-population.  The total number of statistically-independent objects has expectation value $\langle N_{\rm mix}\rangle=\iint\Sigma_{\rm mix}(\vecr)\,\deriv\vecr=\sum_{i=1}^{N_{\rm pop}}\langle N_i\rangle$, where $\langle N_i\rangle$ is the expectation value for the number of independent objects in the $i^{\rm th}$ sub-population.  The total surface number density of countable items is $\Sigma'_{\rm mix}(\vecr)=\sum_{i=1}^{N_{\rm pop}}\Sigma'_i(\vecr)=\sum_{i=1}^{N_{\rm pop}}\bigl (1+f_{{\rm b}_i}(\vecr)\bigr )\,\Sigma_i(\vecr)$, where $f_{{\rm b}_i}(\vecr)=\Sigma_{{\rm b}_i}(\vecr)/\Sigma_i(\vecr) $ is the local binary fraction of the $i^{\rm th}$ sub-population.  
The total number of countable items has expectation value $\langle N'_{\rm mix}\rangle=\iint\Sigma'_{\rm mix}(\vecr)\,\deriv\vecr=\sum_{i=1}^{N_{\rm pop}}\bigl (1+f_{{\rm b}_i}(\vecr)\bigr )\langle N_i\rangle$.

Generalizing the derivation from Appendix \ref{app:1} to include a mixture of sub-populations, the joint position-separation function for countable items is
\begin{align}
    \psi'_{\rm mix}(s,\vecr)
    \approx\sum_{i=1}^{N_{\rm pop}}\sum_{j=1}^{N_{\rm pop}}\bigl (1+f_{{\rm b}_i}(\vecr)+f_{{\rm b}_j}(\vecr)\nonumber\\
    +f_{{\rm b}_i}(\vecr)\,f_{{\rm b}_j}(\vecr)\bigr )\,\psi_{i,j}(s,\vecr)\nonumber\\
    +2\sum_{i=1}^{N_{\rm pop}}f_{{\rm b}_i}(\vecr)\,\Sigma_i(\vecr)\,\phi_{{\rm b}_i}(s|\vecr),
    \label{eq:psiprime_mix}
\end{align}
where $\phi_{{\rm b}_i}(s|\vecr)$ is the (normalized) internal separation function for binaries at position $\vecr$ within the $i^{\rm th}$ sub-population.  The conditional and marginal separation functions for countable items can then be calculated as $\mu'_{\rm mix}(s|\vecr)=\psi'_{\rm mix}(s,\vecr)/\Sigma'_{\rm mix}(\vecr)$ and $\phi'_{\rm mix}(s)=\iint \psi'_{\rm mix}(s,\vecr)\,\deriv\vecr$, respectively.

If the binary fractions and internal separation functions are all independent of position, then the marginal separation function for countable items is
\begin{align}
    \phi'_{\rm mix}(s)\approx\sum_{i=1}^{N_{\rm pop}}\sum_{j=1}^{N_{\rm pop}}(1+f_{{\rm b}_i}+f_{{\rm b}_j}+f_{{\rm b}_i}f_{{\rm b}_j})\,\phi_{i,j}(s)\nonumber\\
    +2\sum_{i=1}^{N_{\rm pop}}\langle N_i\rangle \,f_{{\rm b}_i}\,\phi_{{\rm b}_i}(s),
\end{align}
and the joint position-separation function and marginal separation function for countable items integrates to
\begin{align}
    \iiint \psi'_{\rm mix}(s,\vecr)\,\deriv s\,\deriv\vecr=\int\phi'_{\rm mix}(s)\,\deriv s\hspace{1in}\nonumber\\
    \approx \sum_{i=1}^{N_{\rm pop}}\sum_{j=1}^{N_{\rm pop}}(1+f_{{\rm b}_i}+f_{{\rm b}_j}+f_{{\rm b}_i}\,f_{{\rm b}_j})\,\langle N_i\rangle\,\langle N_j\rangle\nonumber\\
    +2\sum_{i=1}^{N_{\rm pop}}f_{{\rm b}_i}\,\langle N_i\rangle.
\end{align}

\subsection{Some Useful Cases with Analytic  Results}
We now consider some specific cases in which the integrals in Equations \ref{eq:mu_general} and \ref{eq:phi} can be calculated analytically.  First we note that if $\Sigma(\vecr)$ has circular symmetry about the origin, Equations \ref{eq:mu_general} and \ref{eq:phi} become 
\beq 
    \label{eq:int1} \mu(s|R)=\int_0^{2\pi} \Sigma(\sqrt{R^2 + s^2 + 2Rs\cos{\alpha}}) s \,\mathrm{d}\alpha 
\eeq
and
\beq 
    \label{eq:int2} 
    \phi(s) = \int_0^{2\pi} \int_0^{\infty} \Sigma(R) \,\mu(s|R)\, R\, \mathrm{d}R\, \mathrm{d}\theta,
\eeq
respectively.   

\subsubsection{Uniform Density in a Circular Field}
Consider a population of objects drawn from a (2D) position function that specifies uniform surface number density $\Sigma_{u,0}$ within a circle of finite radius $R_{\rm max}$, and zero density at $R>R_{\rm max}$:  
\beq 
    \label{eq:UniSpa} \Sigma_u(R) = \Sigma_{u,0} \,\Theta (R_{\rm max} - R),
\eeq 
where $\Theta$ is the Heaviside-Theta function.  The number of objects has expectation value $\langle N_u\rangle=\pi\,R_{\rm max}^2\,\Sigma_{u,0}$.  In terms of dimensionless radius $R_u\equiv R/R_{\rm max}$ and dimensionless separation variable $s_u\equiv s/R_{\rm max}$, the conditional separation function is (via Equation \ref{eq:int1})
\begin{align}
    \notag
    \mu_{\rm u}(s_u|\vecr)=\frac{2\,\langle N_u\rangle \,s_u}{\pi R_{\rm max}}\biggl [\pi \Theta\bigl (R_{\rm max}(1-R_u-s_u)\bigr ) + \\
    \biggl(\pi - \sec^{-1}\left(\frac{-2R_u s_u}{R_u^2+s_u^2-1}\right) \Theta^{\prime}\biggr )\biggr ],
    \label{eq:uniform_conditional}
\end{align}
where 
\begin{equation*}
    \Theta^{\prime} \equiv \bigl (-1 +\Theta\bigl (R_{\rm max}[1-R_u-s_u]\bigr )\bigl (-1+\Theta\bigl (R_{\rm max}[s_u-R_u-1]\bigr )\bigr ).
\end{equation*}
The marginal separation function can be derived via multiple methods (\cite{hammersley1950distribution}, \cite{lellouche2020distribution}), and is given by
  \beq \label{eq:UniSep} \phi_{u}(s_u) = \frac{4\,\langle N_u\rangle ^2\,s_u}{\pi R_{\rm max}} \left[ \cos^{-1}\left(\frac{s_u}{2}\right) - \frac{s_u}{2}\sqrt{1-\frac{s_u^2}{4}}\right] \eeq
  for $s_u \in [0,2]$, and $\phi_{u}(s_u)=0$ at $s_u>2$.
\begin{figure}
  \includegraphics[width=0.4\textwidth]{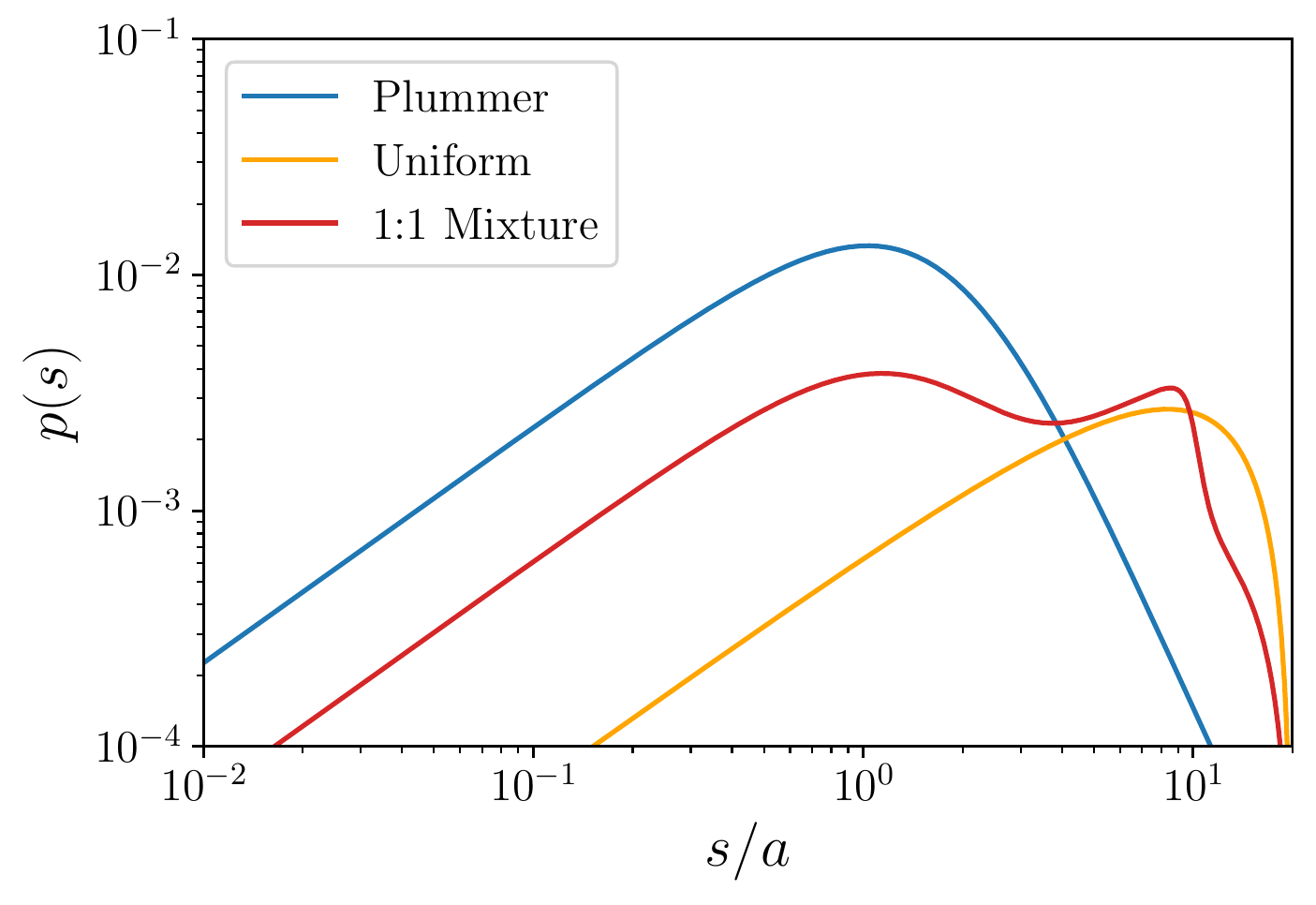}
  \caption{Marginal probability density for pair separations within 1) a Plummer sphere of projected halflight radius $a$ (blue), 2) objects distributed uniformly within a circle of radius $R_{\rm max}=10a$ (orange), and 3) a 1:1 mixture of the two (red), for separations up to $2R_{\rm max}$.}
  \label{fig:AllSep}
\end{figure}

\begin{figure*}
  \centering
  \includegraphics[width=\linewidth]{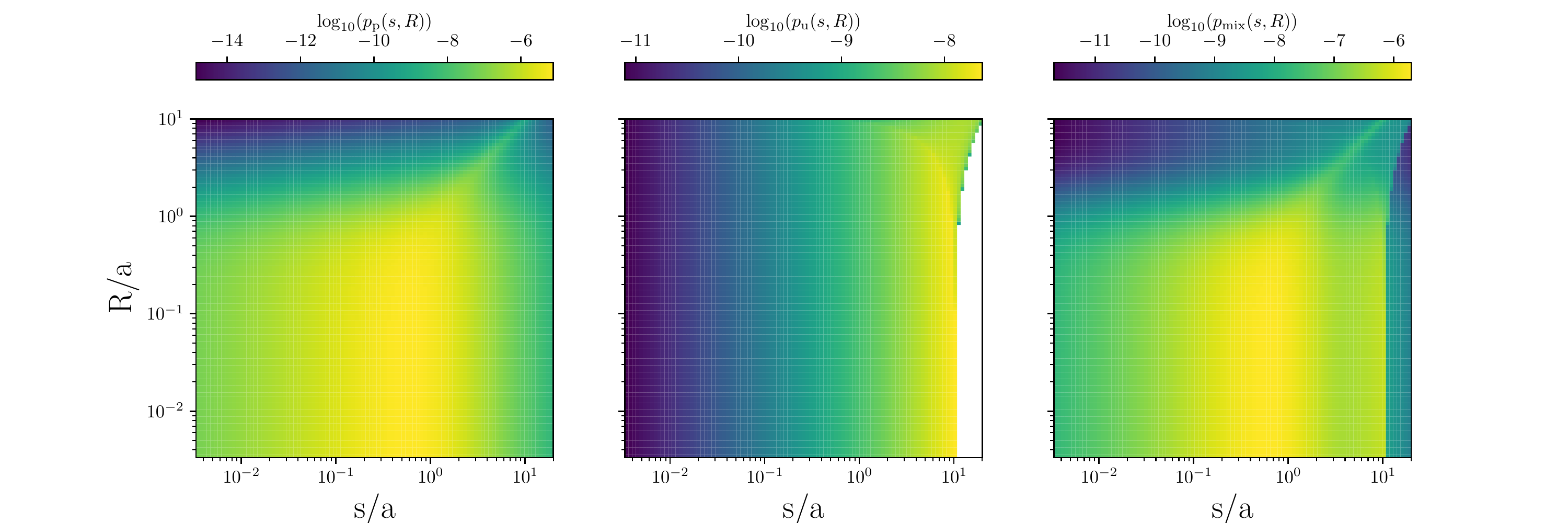}  
\caption{Joint probability density of object positions and pair separations, for (left to right): a Plummer sphere of projected halflight radius $a$, objects distributed uniformly within a circle of radius $R_{\rm max}=10a$, and a 1:1 mixture of the two, for separations up to $2R_{\rm max}$.  In the center panel, whitespace corresponds to regions where $p(s,\vecr)=0$ due to the finite field size.}
\label{fig:jointpsr}
\end{figure*}

\begin{figure*}
  \centering
  \includegraphics[width=\linewidth]{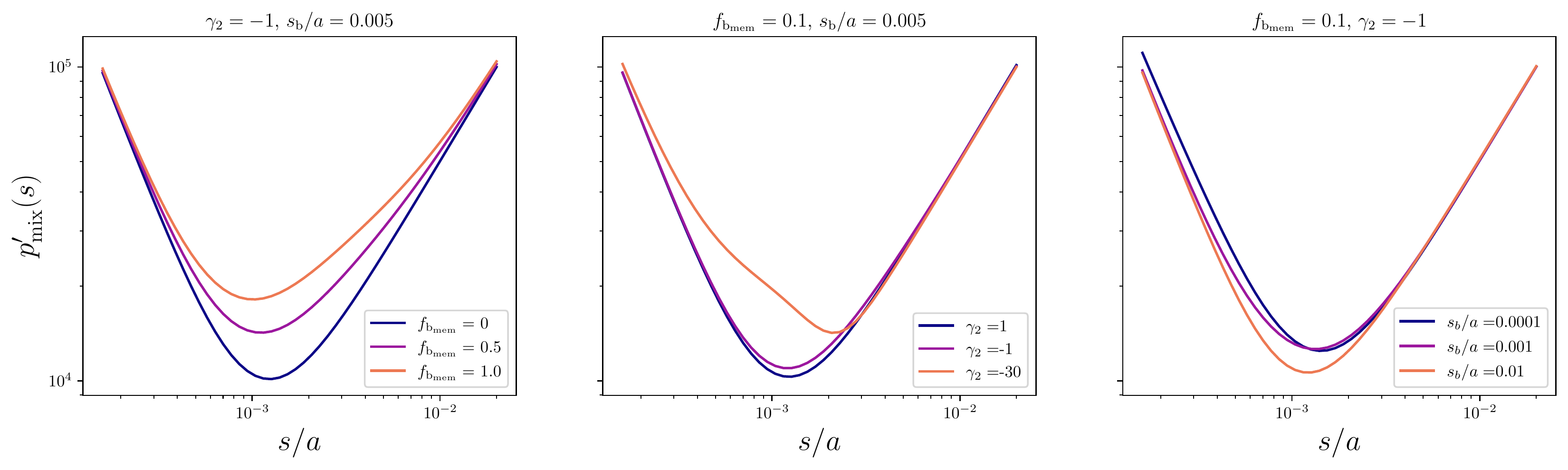}  
\caption{\scriptsize Marginal probability density of pair separations for a 1:1 mixture of Plummer and uniform sub-populations, each containing binary pairs.  The separation function for binary pairs within the uniform sub-population (with binary fraction 0.1) follows a fixed broken power law (Equation \ref{eq:BPL}), with indices $\gamma_1=-1.55$ and $\gamma_2=-3.33$, break separation $s_{\rm b}= 0.001\,a$ and smoothing parameter $\Lambda=0.67$.  Binary pairs within the Plummer sub-population follow a broken power law with $\gamma_1=+0.5$, $\Lambda=0.5$ and the other parameters allowed to vary.  Left, center, right panels show how the marginal density of pair separations within the mixture varies with binary fraction, power-law index $\gamma_2$ and break separation, respectively, of the binaries within the Plummer sub-population.}
\label{fig:vary_params}
\end{figure*}

\subsubsection{Plummer sphere}
The \citet{plummer} model is commonly used to fit the stellar density profiles in dwarf spheroidal galaxies \citep[e.g,][]{mcconnachie2012observed,moskowitz2020stellar}.  Objects within a Plummer sphere have positions drawn from 
  \beq
  \label{eq:PlumSpa}
  \Sigma_p(R)=\frac{\Sigma_{p,0}}{(1+R^2/a^2)^2},
  \eeq
where the number of stars has expectation value $\langle N_p\rangle=\pi a^2\Sigma_{p,0}$ and half the stars are expected to be enclosed within a circle of radius $a$.  Via Eq. \ref{eq:int1}, and in terms of dimensionless separation variable $s_p\equiv s/a$, the conditional separation function is
\beq
    \mu_p(s_p|\vecr)=\frac{2\,\langle N_p\rangle\,(1+s_p^2+R^2/a^2)\,s_p/a}{\bigl [(1+(R/a-s_p)^2)(1+(R/a+s_p)^2)\bigr ]^{3/2}}.
    \label{eq:plummer_conditional}
\eeq
Via Eq. \ref{eq:int2}, the marginal separation function is
  \begin{eqnarray} \label{eq:PlumSep}
    \phi_{p}(s_p) = \frac{4\,\langle N_p\rangle^2\,s_p/a}{(4\,s_p+s_p^3)^3} [-8\,s_p+2\,s_p^3+s_p^5  \nonumber \\
    + 2\,(1+s_p^2)\sqrt{4+s_p^2}\operatorname{tanh}^{-1}(b) ],
\end{eqnarray}
where $b \equiv \frac{s_p(2+s_p^2)\sqrt{4+s_p^2}}{2+4s_p^2 + s_p^4}$.

\subsubsection{Mixture of Plummer Sphere and Uniform Background} Over the few square degrees (or less) subtended by most dwarf galaxies, stars in the Galactic foreground follow approximately uniform distributions \citep{ih95}.  Therefore most nearby dSphs can be modeled as mixtures of $N_{\rm plum}$ stars that follow a Plummer distribution with Plummer radius $a$, and $N_{\rm u}$ stars that follow a uniform distribution over a circular field of radius $R_{\rm max}$.  

In that case, the conditional separation function, $\mu_{\rm mix}(s|\vecr)$, is given by Equation \ref{eq:mix_conditional}, with individual terms for uniform and Plummer sub-populations specified by Equations \ref{eq:uniform_conditional} and \ref{eq:plummer_conditional}, respectively.  The marginal separation function, $\phi_{\rm mix}(s)$, is given by Equation \ref{eq:mix_marginal}, with analytic contributions from cross terms (see Equation \ref{eq:mix_marginal}) $\phi_{u,u}(s_u)$, $\phi_{p,p}(s_p)$ and, in terms of dimensionless separation variable $s_p\equiv s/a$, 
\beq 
    \label{eq:PlumUniCross}
    \phi_{u,p}(s_p) = \phi_{p,u}(s_p) = \frac{N_u\,N_p\,s_p \left(z^2-s_p^2+ \sqrt{c}-1\right)}{zR_{\rm max}\sqrt{c}},
\eeq  
where $z\equiv R_{\rm max}/a$ and $c\equiv 1+z^2+2s_p^2(1-z^2)+s_p^4$.

Figure \ref{fig:AllSep} displays marginal probability densities for separations, $p(s)=\phi(s)/\bigl (\int\phi(s)\,\deriv s\bigr )$, for populations of objects that follow a Plummer profile (blue) with scale radius $a$, and a uniform distribution (orange) within a field of radius $R_{\rm max}=10a$.  Both curves scale as $p(s)\propto s$ at small separation, reaching maxima at separations near the relevant scale radii of $a$ and $R_{\rm max}$, respectively.  The red curve indicates the marginal probability density, $p_{\rm mix}(s)=\phi_{\rm mix}(s)/\bigl (\int\phi_{\rm mix}(s)\,\deriv s\bigr )$, for a 1:1 mixture of the Plummer and uniform populations, showing local maxima at both characteristic scales.  

For the same three cases, heatmaps  in Figure \ref{fig:jointpsr} display joint probability densities $p(s,\vecr)=\psi(s,\vecr)/\bigl (\iiint \psi(s,\vecr)\,\deriv s\,\deriv\vecr\bigr )$ and $p_{\rm mix}(s,\vecr)=\psi_{\rm mix}(s,\vecr)/\bigl (\iiint\psi_{\rm mix}(s,\vecr)\,\deriv s\,\deriv\vecr \bigr )$.  Separations within the Plummer sphere correlate strongly with position (left panel)---a consequence of the fact that the  probability density of radial coordinate $R$ approaches zero as $R\rightarrow \infty$.  Thus the rare star at large radius in the Plummer sphere will tend to be widely separated from almost all other stars, which tend to reside at smaller radius.  In contrast, the uniform distribution displays no obvious correlation, with the separation density peaking near $s\approx R_{\rm max}$ except near $R\approx R_{\rm max}$, where the most probable separation decreases slightly due to the field-edge effect (middle panel).

Figure \ref{fig:vary_params} shows what happens to the marginal probability densities if we add binary components to both uniform and Plummer sub-populations within the 1:1 mixture.  In this case the binary systems within both sub-populations follow a broken power law separation function (Equation \ref{eq:BPL}).  Parameters of the binary separation function within the uniform sub-population are held fixed at values chosen to represent observational constraints on wide binaries within the Milky Way halo \citep{tian2019separation}, with power-law indices $\gamma_1=-1.55$, $\gamma_2=-3.33$, break separation $s_{\rm b}=0.001\, a$ , and smoothing parameter $\Lambda=0.67$.  Panels in Figure \ref{fig:vary_params} then show the effect of varying the binary fraction (left), `outer' power-law index $\gamma_2$ (middle), and break separation (right) of the binary separation function within the Plummer sub-population.  In all cases, the marginal probability density, $p'_{\rm mix}(s)=\phi'_{\rm mix}(s)/\bigl (\int\phi'_{\rm mix}(s)\,\deriv s\bigr )$, is now characterized generally by a transition from domination by binaries within the uniform distribution at small separation (where $\phi_{{\rm b}_{\rm u}}(s)\propto s^{-1.55}$) to domination by physically unassociated pairs at large separation.  Details of the transition change with the binary fraction and separation function that we adopt for the Plummer sub-population.  As the binary fraction within the Plummer sub-population increases (left panel), the minimum in the $p'_{\rm mix}(s)$ function shifts toward smaller $s$, a result of the fact that the separation function for the binaries within the Plummer sub-population peaks at finite separation $s\sim a$.  As the `outer' index $\gamma_2$ changes from positive to strongly negative (middle panel), binaries with large separation become scarce within the Plummer sub-population, and the $p'_{\rm mix}(s)$ curve increases toward smaller $s$.  Finally, as the break separation increases (right panel), the minimum in the $p'_{\rm mix}(s)$ function again shifts toward smaller $s$ as a larger fraction of binaries within the Plummer sub-population have separations $s<s_b$.  We emphasize that all of these behaviors can change in detail depending on the binary separation functions assumed for both sub-populations; our purpose here is merely to provide an example of how the observable marginal density $p'_{\rm mix}(s)$ can be sensitive to the binary separation function that we seek to infer.

\section{Application to Dwarf Galaxies} \label{sec:Application}

We now apply this formalism to investigate the potential for detectability  and characterization of wide binary systems within nearby dwarf galaxies.  We generate and analyze mock observational catalogs designed to reproduce the observed structural parameters of the 40 dSph galaxies analyzed by \citet{moskowitz2020stellar}, with binary companions inserted by hand according to assumed separation functions.  The galaxy sample spans a range in luminosity of $10^3\la L_V/{L_{V,\odot}}\la 10^7$, Plummer radius $10^1\la a/\mathrm{pc}\la 10^3$, and distance $20\la D/\mathrm{kpc}\la 250$, providing a natural `grid' for studying the dependence of detection sensitivity on intrinsic properties. 

\subsection{Generation of Mock Data}
\label{sec:generation}

For a given dwarf galaxy, we adopt published values for luminosity, distance and metallicity from the review of \citet{mcconnachie2012observed} or, when necessary, from more recent discovery papers \citep[e.g.,][]{koposov2015beasts,drlica2015eight}.  We adopt `circularized' Plummer radii fit by \citet[][fourth column of their Table 3]{moskowitz2020stellar}.   Given the adopted metallicity and assuming old (12 Gyr) age, we use PARSEC isochrones and luminosity functions \citep[][assuming the default \citet{kroupa01,kroupa02} initial mass function]{bressan2012parsec} to sample the present-day luminosity function.  

We assume that the dSph member sub-population intrinsically consists of $N_{\rm p_{mem}}$  `parents' that have positions distributed according to a circular Plummer profile (Equation \ref{eq:PlumSpa}), and $N_{\rm b_{mem}}$ `binary companions' to a fraction $f_{\rm b_{mem}}$ of the parents.  We assume that the luminosities of binary companions are drawn independently from the same luminosity function as those of the parent population.  We assume that the binary fraction and binary separation function within the dSph member sub-population are both independent of the parent's position and luminosity, with the separation function,  $\phi_{\rm mem}(s)$, following one of three distinct functional forms, discussed in detail in Section  \ref{sec:BinaryModels}. 

We assume that the dwarf galaxy is observed against a foreground sub-population of Milky Way stars that consists of parents following a uniform spatial distribution within a circular field of radius $R_{\rm max}=10$ times the Plummer radius of the dwarf galaxy, and binary companions to a fraction $f_{\rm b_{non}}$ of those parents.  We assume that, within this field, the number of detected  (i.e., brighter than the adopted magnitude limit) parents within the nonmember sub-population equals the number of detected parents within the dSph member sub-population.  We assume that the binary fraction and binary separation function within the nonmember sub-population are both independent of position, with the separation function, $\phi_{\rm non}(s)$, following an observationally-motivated  broken power law  (Section \ref{sec:BinaryModels}). 

We construct a mock observational catalog as follows:
\begin{enumerate}

    \item Draw $N_{\rm mem}$ luminosities randomly from the present-day stellar luminosity function function.  The value of $N_{\rm mem}$, the number of stars belonging to the dwarf galaxy (including parents and binary companions), is set by the requirement that the cumulative luminosity equal the published galaxy luminosity.  
    
    \item Given the adopted member binary fraction, randomly assign  each member star the status of either `parent' or `binary companion'. 
    
    \item For each parent in the member sub-population, assign a 2D position by sampling radial coordinate $R$ from a Plummer distribution having the published Plummer radius, and position angle $\theta$ from a uniform distribution between $0\leq\theta\leq 2\pi$. 
    
    \item For each binary companion in the member sub-population, assign a 2D position by offsetting from a randomly-chosen (without replacement) parent by a 2D vector $\Delta\vecr$.  Draw the magnitude $|\Delta\vecr|=s$ from the adopted member binary separation function (Section \ref{sec:BinaryModels}), and the direction angle from a uniform distribution between $0\leq \alpha\leq 2\pi$.  
    
    \item Impose observational resolution and magnitude limits by 1) combining into a single point source any pairs  separated by less than an assumed resolution limit $r_{\rm lim}$, and 2) discarding any source fainter than an assumed limiting magnitude $m_{\rm lim}$.  
        
    \item Add foreground nonmembers by repeating step 3, but for $N_{\rm p_{non}}$ non-member parents (all of which are assumed to be brighter than the adopted magnitude limit) with 2D positions drawn from a uniform spatial distribution over a circle of radius $R_{\rm max}=10$ times the dSph Plummer radius.  Then, to a fraction $f_{\rm b_{non}}$ of the nonmember parents (drawn randomly without replacement), add a binary companion with position offset as described in step 4, drawing the magnitude of the offset from  $\phi_{\rm non}(s)$.  Since the separation function is expressed in physical units (Section \ref{sec:BinaryModels}), convert to angular offsets (assuming the sky is flat over the observed field) by assuming a characteristic nonmember distance of 10 kpc.
    
    \item In order to mimic observational errors in the measurement of centroids, scatter the position of each point source by a 2D vector drawn randomly from a bivariate Gaussian that is centered on the true location and has covariance matrix given by $0.5\, r_{\rm lim}\, I_2$, where $I_2$ is the $2\times2$ identity matrix.  
\end{enumerate}

We adopt fiducial values of $V_{\rm lim}=27$ and $r_{\rm lim}=0.05$ arcsec for the magnitude and resolution limits, respectively.  For the member sub-population, we adopt a binary fraction such that a fraction $f_{\rm b_{mem}}=0.1$ of member parents have a binary companion with separation $s<5$ pc that is drawn from the initial broken power-law model (without the imposed upper limit, the total number of separations diverges when $\gamma_2\geq -1$).  This fraction then effectively decreases for the truncated power-law input and increases for the {\"O}pik's law input, as discussed in Section \ref{sec:BinaryModels}. 

The empirically-motivated separation function that we adopt for nonmembers is a broken power law that diverges toward $s\rightarrow 0$ (Section 3.2).  Therefore we choose the nonmember binary fraction such that the number of nonmember binary companions that are separated from their parents by more than the adopted resolution limit equals $0.1$ times the number of nonmember parents.

After imposing these limits, the number of `detectable' binary systems within both member and non-member sub-population---i.e., pairs separated by more than the adopted resolution limit and for which both members are brighter than the adopted magnitude limit---ranges from a few to $\sim 10^4$.  

We reiterate that the wide binary fraction within dwarf galaxies is currently unconstrained observationally.   While \citet{longhitano10} infer a wide binary fraction of $\sim 10\%$ within the solar neighborhood, this constraint is not necessarily relevant for the older, more metal-poor stellar populations that occupy the denser dark matter halos of dwarf galaxies.  Our assumption that $f_{\rm b_{mem}}=0.1$ for the broken power law separation function is meant only to provide---given the range of luminosities and distances amongst the known dwarf galaxies---a corresponding range in detectability of their wide binary populations.  While the mock wide binary populations provide some basis for comparing relative detectability across different dwarf galaxies, and for testing performance as a function of number of detectable wide binaries, they should not be interpreted as forecasts of wide binary populations---either in individual dSphs or in the dSph population.  

\subsection{Input Binary Separation Functions} \label{sec:BinaryModels}
Separations between companions within binary star systems are set by the physics of star formation/evolution and interactions with the ambient medium \citep{chandra44,heggie75,bahcall85,weinberg87,kouwenhoven10,jiang10,penarrubia20}.  As such, binary separations can be independent of the density field, $\Sigma(\vecr)$, of the parent stellar population---at least on scales smaller than the characteristic parent separation---and the  binary separation function can be modeled directly using simple analytic formulae.  

For example, \citet{opik24} proposes a log-uniform binary separation function, $\phi(s)\propto s^{-1}$, which is typically observed within young star clusters and stellar associations \citep{kouwenhoven07,kraus08}.  Working primarily with short-period binary systems within the solar neighborhood,  \citet{duquennoy1991multiplicity} and \citet{raghavan10} fit log-normal period distributions that, assuming uniform distributions of eccentricity and mass ratio, correspond  approximately to log-normal separation functions.  In order to allow for a characteristic scale for wide binary formation/destruction, others have adopted a broken power law  \citep[e.g.,][]{andrews17,el-badry18,tian2019separation}:
\beq \label{eq:BPL}
\phi(s) =\phi_0\, \biggl (\frac{s}{s_b}\biggr)^{\gamma_1} \left[  \left( 1 +\bigl (\frac{s}{s_b}\bigr )^{\frac{1}{\Lambda}} \right) \right]^{(\gamma_2-\gamma_1)\Lambda},
\eeq
such that near a `break' separation $s_b$, the power-law index changes from `inner' value $\gamma_1$ to `outer' value $\gamma_2$, at a rate controlled by smoothing parameter $\Lambda$.

For the purpose of testing our methodology, we generate mock data sets for which the wide binary populations follow theoretically and/or observationally motivated separation functions. 

We glean theoretical motivation from the work of \citet[][`P21' hereafter]{penarrubia2021creation}, who conducts N-body experiments to study the formation of wide binaries via dynamical capture within the tidal debris of disrupting star clusters.  While this formation process is stochastic, the resulting pairs follow a universal semi-major axis distribution that scales as $p(a)\propto a^{1/2}$, with normalization that scales with progenitor cluster mass as $N_{\rm b}\propto M^{1/2}$. Subsequently, perturbative encounters with `clumpy' substructures (e.g., dark matter subhalos) cause the separation function to evolve toward {\"O}pik's law beyond a characteristic separation that decreases with time.  Simultaneously, beyond another characteristic separation that depends on the smooth component of the host's gravitational potential, tidal forces steepen the separation function more dramatically, over time approaching a sharp truncation.

In order to gauge our ability to detect and characterize dSph wide binary populations under some combination of these processes,  for each dSph we generate three separate mock data sets, each using a unique input function to draw binary separations for the dSph members.  In each case the input function follows a broken power law of the form given by Equation \ref{eq:BPL}, but with parameters chosen to represent different combinations of the processes simulated by P21.   

\begin{itemize}
    \item \textit{Broken power law (BPL):} The first input separation function follows a broken power law with index changing smoothly ($\Lambda=0.5$) from $\gamma_1=+0.5$ to $\gamma_2=-1$ around break break separation $s_b=0.5$ pc.  Following P21, this behavior represents a wide binary population that forms via dynamical capture and evolves only mildly due to encounters with perturbers (cf. right panel of P21's Figure 8).  

    As stated in Section \ref{sec:generation}, for the BPL model we adopt a member binary fraction of $f_{\rm b_{mem}}=0.1$; the fraction of member parents that have \textit{detectable} binary companions (i.e., separated by more than the adopted resolution limit) is smaller by an amount that depends on distance to the dSph.  

    \item \textit{Truncation:} The second input separation function follows a broken power law with index changing sharply ($\Lambda=0.01$) from $\gamma_1=+0.5$ to $\gamma_2=-\infty$ around $s_b=0.5$ pc.  This behavior represents the truncation that results from tidal disruption of wide binaries in the smooth potential of the host system (cf. left panel of P21's Figure 8).
    
    In practice, we generate realizations of the truncation model simply by removing from the BPL case any binary companions that are  separated from their parents by $s>0.5$ pc. 

    \item \textit{{\"O}pik's law:} The third input separation function is {\"O}pik's law, which we recover by setting $\gamma_1=\gamma_2=-1$.  Physically, this case can represent a scenario in which the wide binary population has evolved strongly due to encounters with perturbers, such that the break separation has shrunk to a scale much smaller than the instrumental resolution limit.  
    
    We generate realizations of {\"O}pik's law by drawing separations from a probability distribution $\phi_{\rm mem}(s)\propto s^{-1}$, subject to the constraint that the number of separations at $s>0.5$ pc be unchanged with respect to the original BPL sample.
    
\end{itemize}

Finally, to the sub-population that represents contaminating Milky Way foreground, we assign binary separations motivated by observational constraints.  Specifically, we assume that at wide separations, foreground binaries follow the broken power law that characterizes `halo-like' binaries in the analysis of \textit{Gaia} data by \citet{tian2019separation}:
$(\gamma_1,\gamma_2,\log(s_b/AU),\Lambda) = (-1.55,-3.33,4.59,0.67)$.

\subsection{Fitting Method} \label{sec:FitMethod}

For each mock dSph, we use Bayesian inference to estimate the posterior probability density in the space defined by parameter vector $\vec{\theta}$, where $\vec{\theta}$ specifies the surface density profiles and binary separation functions of `member' (i.e., dSph) and `nonmember' (Galactic foreground) sub-populations.  Given data vector $\vec{D}$, the posterior probability is 
\beq
    p(\vec{\theta}|\vec{D})=\frac{p(\vec{D}|\vec{\theta})\,p(\vec{\theta})}{p(\vec{D})},
\eeq
where $p(\vec{D}|\vec{\theta})$ is the likelihood of the data given the model specified by $\vec{\theta}$, $p(\vec{\theta})$ is the prior probability distribution, and $p(\vec{D})=\int_{\vec{\theta}}\, p(\vec{D}|\vec{\theta})\,p(\vec{\theta})\,\mathrm{d}\vec{\theta}$ is the marginal likelihood.

Supposing the observational data consists of catalogued positions for $N$ detected stars, we consider the data vector $\vec{D}$ that comprises all off-diagonal elements of the $N\times N$ matrix whose $i,j$  element is given by $(\vecr_i,s_{i,j})$, where $s_{i,j}\equiv |\vecr_i-\vecr_j|$.  Neglecting observational errors and covariance amongst the non-independent discrete data points in $\vec{D}$, we approximate the likelihood function as\footnote{We expect stellar centroid  errors to be smaller than $s_{\rm min}$ for stars detected at sufficient signal-to-noise ratio to be included in observational catalogs.  Covariance, however, is necessarily present amongst the discrete data points in $\vec{D}$ as separations between pairs of objects are not independent; our test results include any errors introduced by neglecting this effect.} 
\beq
    p(\vec{D}|\vec{\theta})\approx\prod_{i\neq j}^{N} \frac{\psi'_{\rm mix}(s_{i,j},\vecr_i|\vec{\theta})\,\psi'_{\rm mix}(s_{j,i},\vecr_j|\vec{\theta})}{\bigl (\iint_{\rm field}\int_{s_{\rm min}}^{s_{\rm max}} \psi'_{\rm mix}(s,\vecr|\vec{\theta})\,\mathrm{d}s\,\mathrm{d}^2\vecr\bigr )^2},
    \label{eq:likelihood}
\eeq
where $\psi'_{\rm mix}(s,\vecr|\vec{\theta})$ is given by Equation \ref{eq:psiprime_mix}.  Limits in the normalizing integral can be adjusted to account for  observational selection imposed, e.g., by resolution limits and finite survey area.   Here we set $s_{\rm min}$ equal to the assumed angular resolution limit of 0.05 arcsec, and set $s_{\rm max}$ equal to the angle corresponding to a physical separation of 2 pc at the distance to the dSph (pairs with larger separations are not included when evaluating Equation \ref{eq:likelihood}).  Imposing such a finite upper limit is not strictly necessary, but can make more efficient use of computational  resources, as large separations contain little information about the binary population.  

When fitting the data set, we assume the dSph member sub-population follows a Plummer profile with binary separation function given by the broken power law of  Equation \ref{eq:BPL}; in order to allow the outer index $\gamma_2$ to reach very negative values (mimicking truncation) without unduly skewing its prior, we fit instead a re-scaled parameter,  $\gamma_2'\equiv -\gamma_2/(\gamma_2-2)$, with uniform prior between $-1\leq \gamma_2'\leq +1$.  For the non-member sub-population, we assume a uniform 2D spatial distribution and a binary separation function that also follows the broken power law form of Equation \ref{eq:BPL}, with smoothing parameter fixed at $\Lambda=0.5$.  Table \ref{table:fitparams} lists the twelve free model parameters and identifies boundaries of the uniform priors that we adopt.
\begin{table*}
\centering
    \caption{Free Parameters and adopted priors}
    \label{table:fitparams}
    \begin{tabular}{c c c c} 
      \hline \hline 
      Free Parameter & Description & Range of Uniform Prior & Equation Reference \\ 
      \hline
      $\log_{10}(a)$ & Plummer Scale Radius (radians) & (-6,-1) & \ref{eq:PlumSpa} \\
      $\log_{10}\langle N'_{\rm mem}\rangle$ & Expectation value for) number of member stars & (-2,6) & \ref{eq:PlumSpa} \\  
      $\log_{10}\langle N'_{\rm non}\rangle$ & Expectation value for number of non-member stars & (-2,6) & \ref{eq:UniSpa} \\ 
      \hline
      &\centering{\underline{\emph{Binary Separation Function: Members}}}  & &\\
      $\log_{10}(f_{\rm b_{mem}})$ & Binary Fraction ($f_{\rm b_{mem}}=N_{\rm b_{mem}}/N_{\rm p_{mem}}$)& (-5,0) & \ref{eq:BPL} \\
      $\gamma_1$ & Inner Power Law Index & (-2,1) & \ref{eq:BPL} \\
      $\gamma_2^{\prime} = \frac{-\gamma_2}{\gamma_2-2}$ & Outer Power Law Index & (-1,1) & \ref{eq:BPL} \\
      $s_{\rm b_{mem}}$ & Break Separation (radians)& (0,$\,2\,\mathrm{pc}/D$) & \ref{eq:BPL} \\
      $\log_{10}(\Lambda)$ & Smoothing Parameter & (-2,0) & \ref{eq:BPL} \\
      \hline
      &\centering{\underline{\emph{Binary Separation Function: Nonmembers}}}  & &\\
      $\log_{10}(f_{\rm b_{non}})$ & Binary Fraction ($f_{\rm b_{non}}=N_{\rm b_{non}}/N_{\rm p_{non}}$)& (-5,0) & \ref{eq:BPL} \\
      $\gamma_1$ & Inner Power Law Index & (-5,-1) & \ref{eq:BPL} \\
      $\gamma_2$ & Outer Power Law Index & (-5,-1) & \ref{eq:BPL} \\
      $s_{\rm b_{non}}$ & Break Separation (radians) & (0,2$\times 10^{-4})$ & \ref{eq:BPL} \\
    \end{tabular}
\end{table*}

We use the software package MultiNest \citep{feroz08,feroz2009multinest}, and in particular its Python wrapper, PyMultiNest (\cite{buchner2014x}, to estimate model parameters.  MultiNest uses a nested sampling algorithm \citep{skilling04} to compute the marginal likelihood, a procedure that also provides random samples drawn from the posterior probability distribution function. 

In practice, we adopt a two-step procedure in which we estimate stellar surface density profile parameters (first three rows of Table \ref{table:fitparams}) separately from the binary separation functions.  In the first step, we fit only the stellar density profile.  Assuming the number of stars observed within an area element is a Poisson random variable, the data set of stellar positions, $D_{\vecr}$, has log-likelihood \citep{richardson11}
\beq
    \ln\, p(\vec{D}_{\vecr}|\vec{\theta})=\sum_{i=1}^N\ln\bigl (\Sigma'_{\rm mix}(\vecr_i|\vec{\theta})\bigr) -\iint_{\rm field}\Sigma'_{\rm mix}(\vecr|\vec{\theta})\,\mathrm{d}^2\vecr,
    \label{eq:likelihood_poisson}
\eeq
where $\Sigma'_{\rm mix}(\vecr|\vec{\theta})=\Sigma'_{\rm mem}(\vecr)+\Sigma'_{\rm non}(\vecr)$ is the surface number density of countable stars at position $\vecr$, given the model specified by $\vec{\theta}$.  Using the likelihood given by Equation \ref{eq:likelihood_poisson}, we run MultiNest to obtain a random sample from the posterior probability distribution for the expectation value for the number of member stars, $\langle N'_{\rm mem}\rangle$, the expectation value for the number of non-member stars, $\langle N'_{\rm non}\rangle$, and the plummer radius, $a$, of the member sub-population. 

In the second step, we run MultiNest using the full likelihood function given by Equation \ref{eq:likelihood}, obtaining random samples from posteriors for parameters that specify binary separation functions (rows 4-12 of Table \ref{table:fitparams}) for member and nonmember sub-populations.  In this second step, for each likelihood evaluation, we draw values of $\langle N'_{\rm mem}\rangle$, $\langle N'_{\rm non}\rangle$ and $a$ randomly from the posterior obtained in step 1. 

\section{Results} \label{sec:Results}
\begin{figure*}[h]

  \centering
  \includegraphics[width=0.8\textwidth]{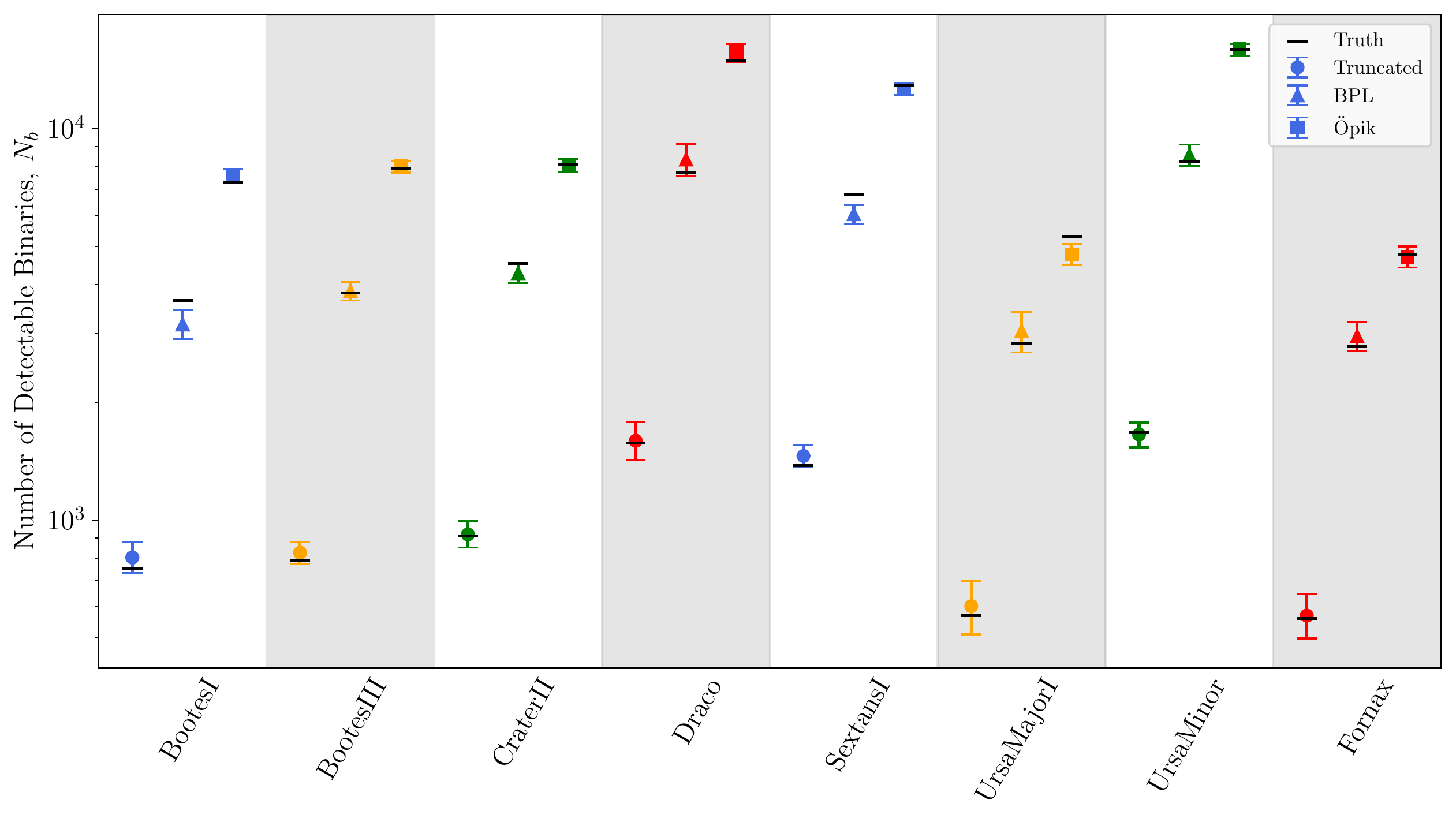}
  \caption{\scriptsize Number of detectable (i.e., separation smaller than the adopted resolution limit, both sources brighter than the adopted magnitude limit) binary systems within the dSph member sub-population, as inferred by applying our analysis to mock data wherein the dSph binary population follows 1) a broken power law, 2) a truncated power law, and 3) {\"O}pik's law, for mock dSphs containing more than 500 detectable binaries.  Data points and errorbars represent median and 95\%  credibility intervals from posterior probability distribution functions; black tick marks identify true input values.  Marker color is intended only to aid the eye in grouping results together for a given dSph.}
  \label{fig:Nb_fits}

\end{figure*}

\begin{figure*}
  \centering
  \includegraphics[width=0.8\textwidth]{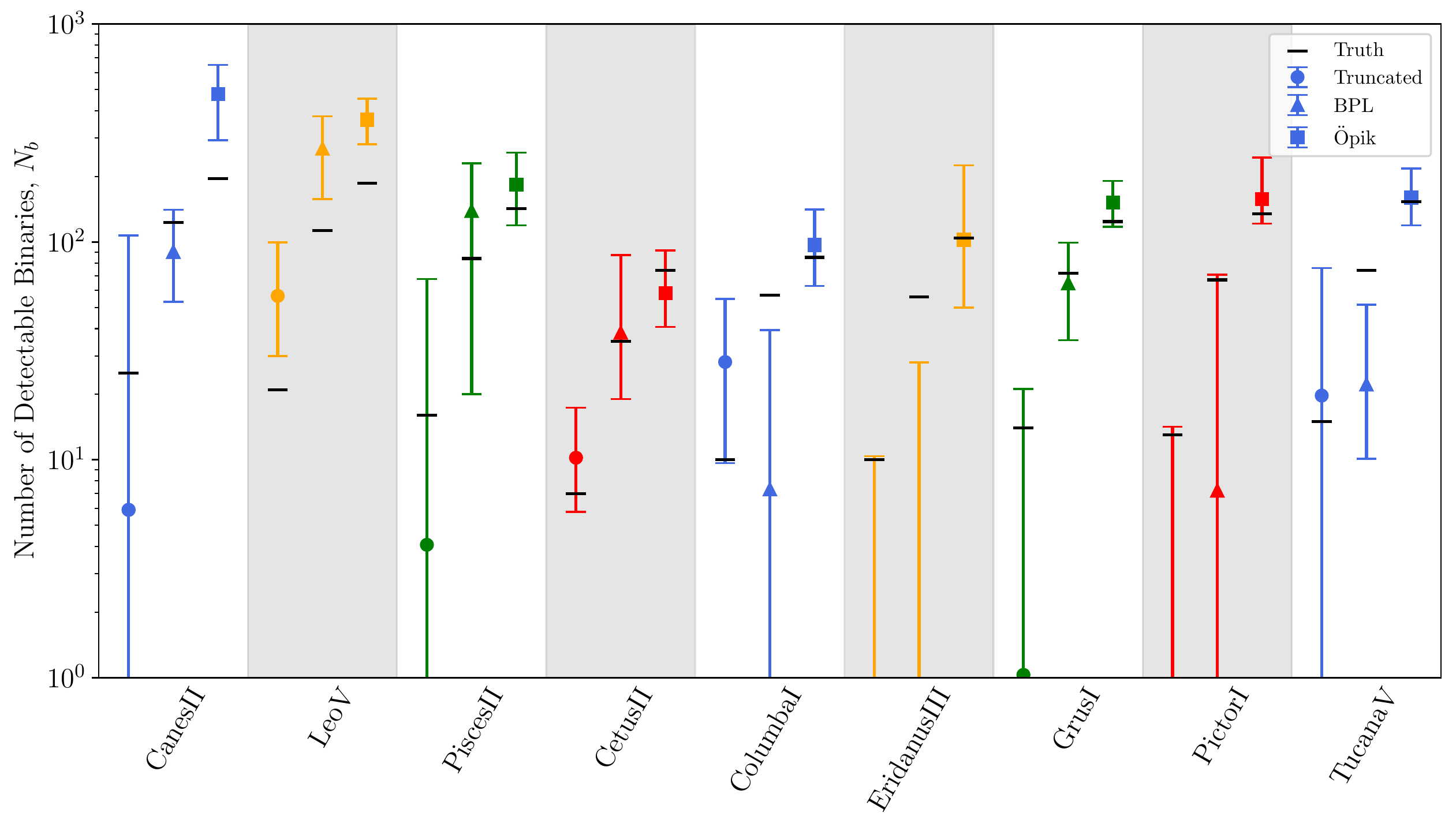}
  \caption{\scriptsize Same as Figure \ref{fig:Nb_fits}, for mock dSphs containing fewer than 200 detectable binary systems.}
  \label{fig:Nb_fits_lt100}
\end{figure*}

\begin{figure*}
  \centering
  \includegraphics[width=0.8\textwidth]{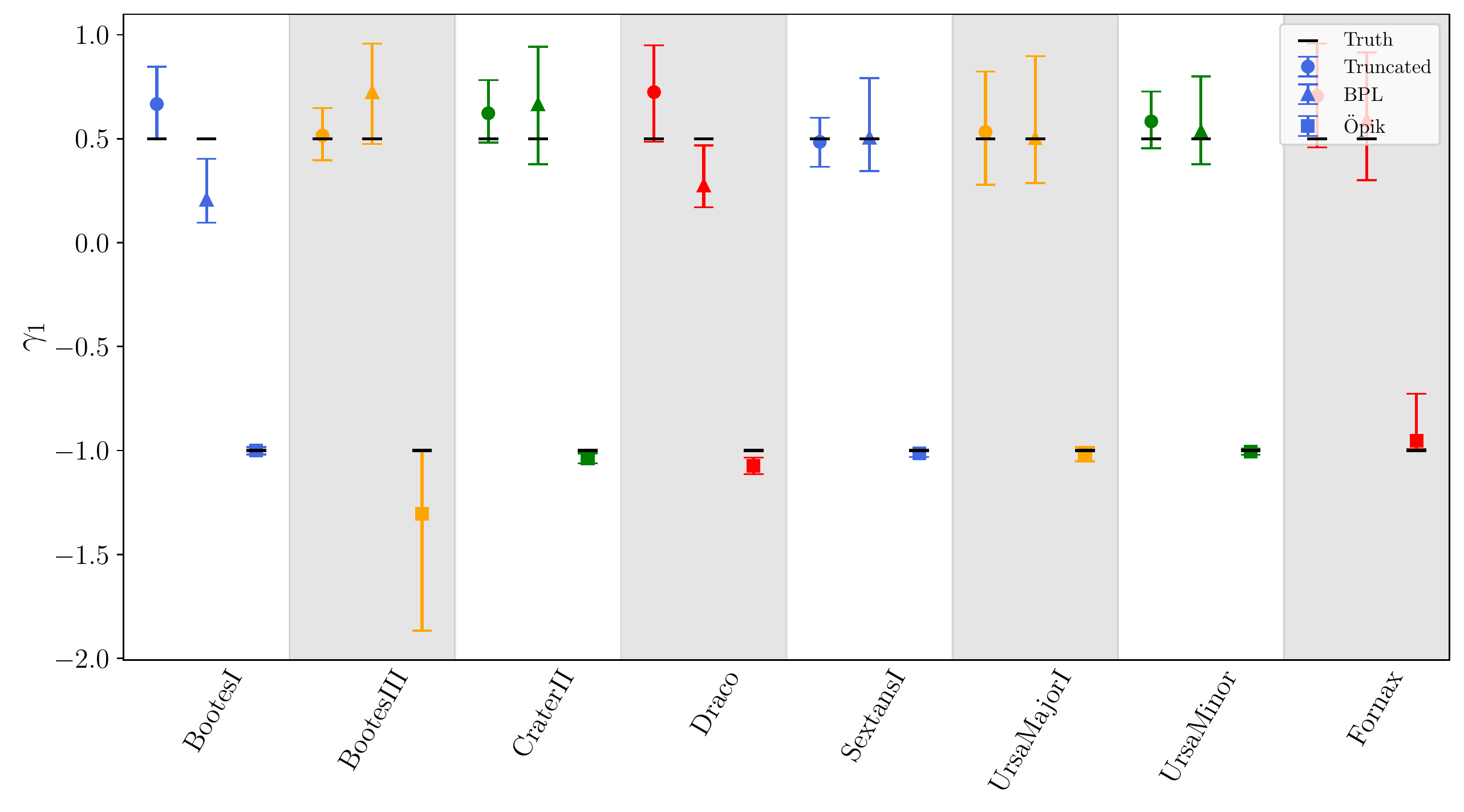}
  \caption{\scriptsize Inference of the inner power law index, $\gamma_1$, of the binary separation function that governs the dSph member sub-population, for dSphs with more than 500 detectable binary systems.  Data points and errorbars represent median and 95\%  credibility intervals from posterior probability distribution functions; black tick marks identify true input values. }
  \label{fig:g1_fits}
\end{figure*}

\begin{figure*}
  \centering
  \includegraphics[width=0.8\textwidth]{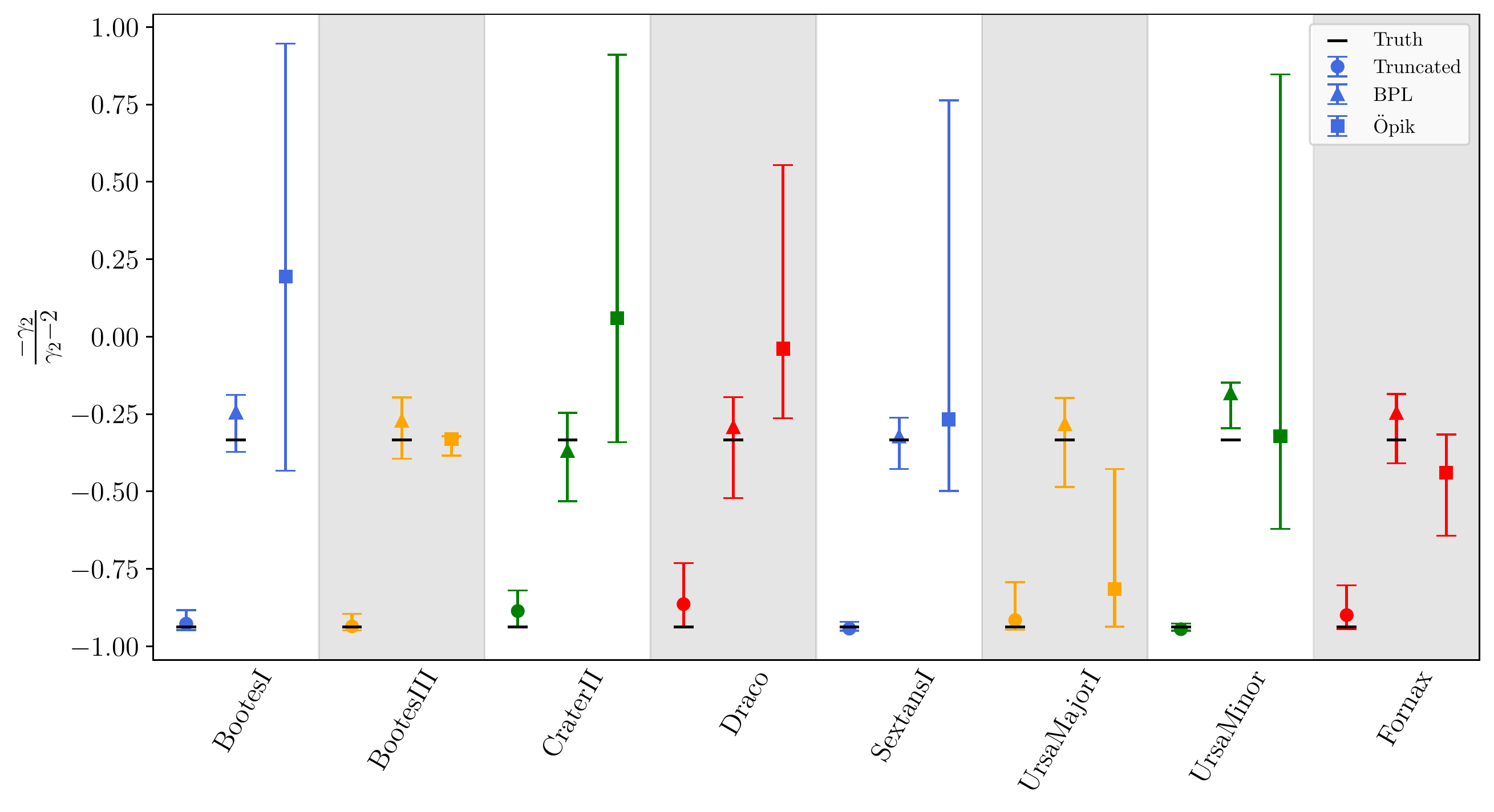}
  \caption{\scriptsize Inference of the outer power law index parameter, $\gamma_2'\equiv -\gamma_2/(\gamma_2-2)$, of the binary separation function that governs the dSph member sub-population, for dSphs with more than 500 detectable binary systems.  Data points and errorbars represent median and 95\%  credibility intervals from posterior probability distribution functions; black tick marks identify true input values (omitted for {\"O}pik's law inputs).}
  \label{fig:g2_fits}
\end{figure*}

\begin{figure*}
  \centering
  \includegraphics[width=0.8\textwidth]{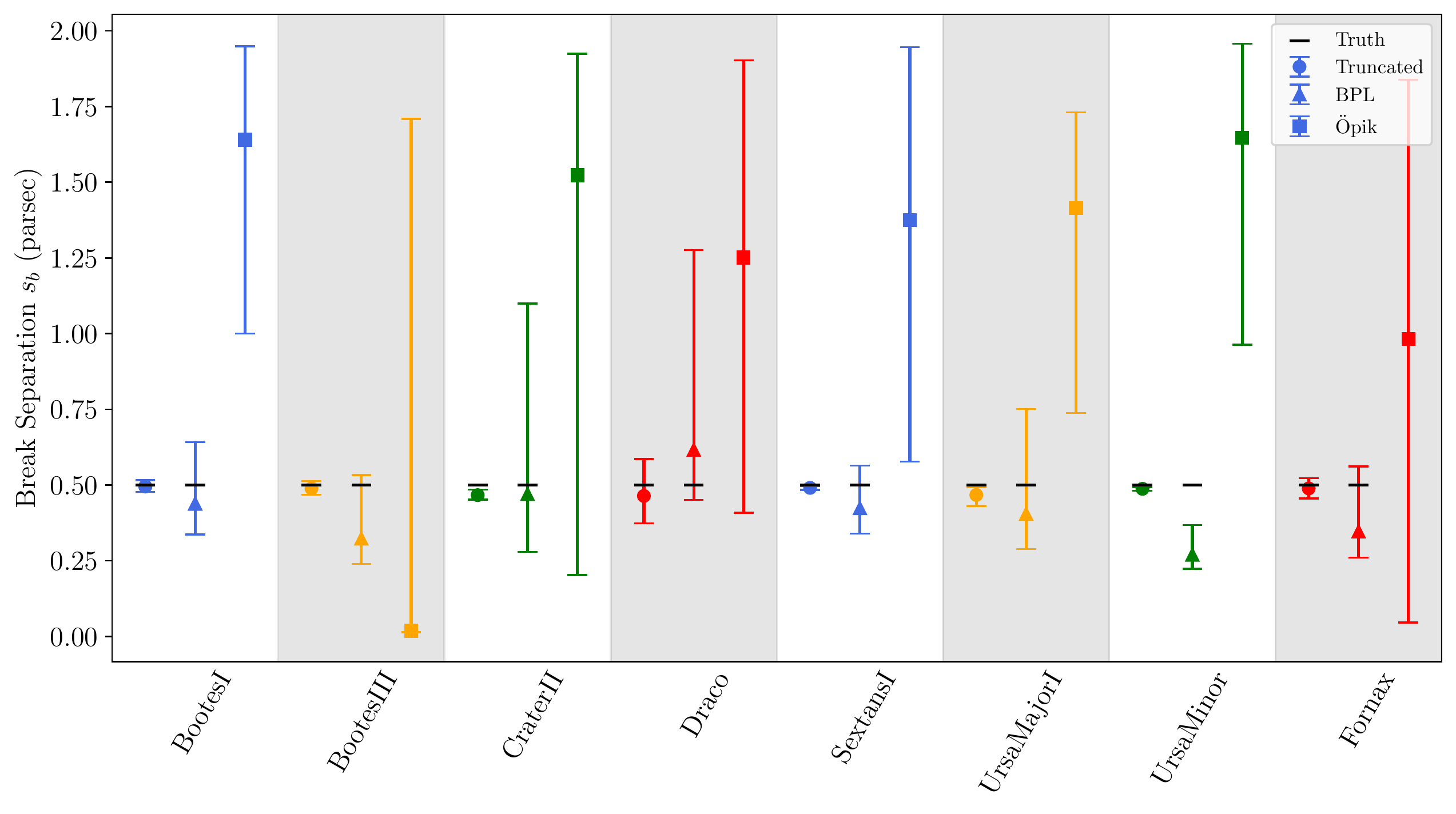}
  \caption{\scriptsize Inference of the break separation, $s_b$, of the binary separation function that governs the dSph member sub-population, for dSphs with more than 500 detectable binary systems.  Data points and errorbars represent median and 95\%  credibility intervals from posterior probability distribution functions; black tick marks identify true input values (omitted for {\"O}pik's law inputs).}
  \label{fig:sb_fits}
\end{figure*}

\begin{figure*}
  \centering
  \includegraphics[width=0.8\textwidth]{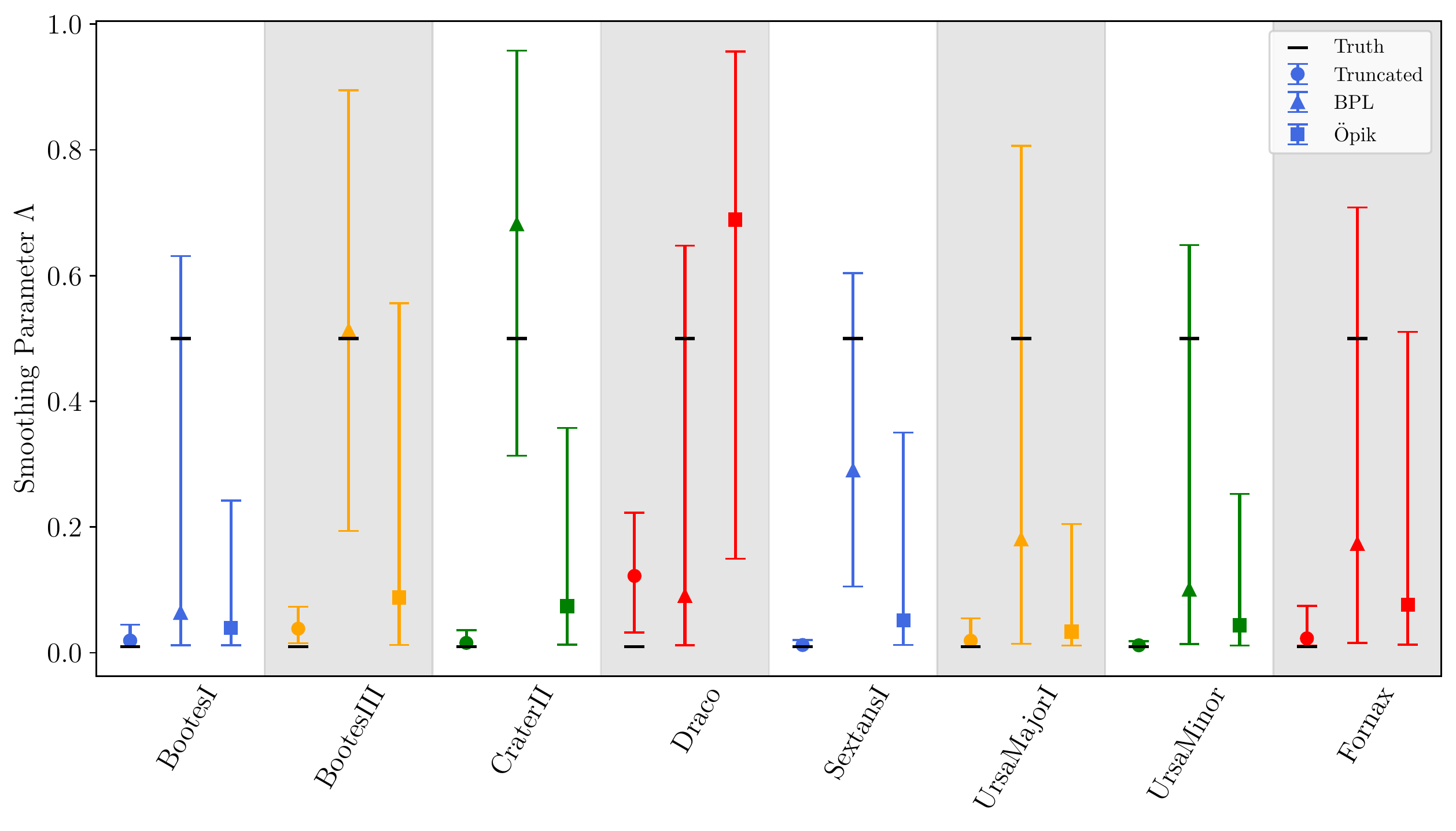}
  \caption{\scriptsize Inference of the smoothing parameter, $\Lambda$, of the binary separation function that governs the dSph member sub-population, for dSphs with more than 500 detectable binary systems.  Data points and errorbars represent median and 95\%  credibility intervals from posterior probability distribution functions; black tick marks identify true input values.}
  \label{fig:L_fits}
\end{figure*}

Here we present results from our analysis of the mock data sets described in Section \ref{sec:Application}.  In general, the initial fits to surface density profiles accurately recover input parameters; we do not display those results here, as our procedure for fitting surface density is already standard practice.  Instead we focus on our inference of binary separation functions for dSph member sub-populations.  Figures \ref{fig:Nb_fits} -    \ref{fig:L_fits} summarize the posterior probability distribution functions that we estimate for parameters that specify the separation function $\phi_{{\rm b}_{\rm mem}}(s)$.  For each of the three input separation functions for each dSph, we plot the median of the posterior probability distribution function inferred for each parameter, with errorbars enclosing the 95\% credibile region.  Black tick-marks indicate true input values.  

Figures \ref{fig:Nb_fits} and \ref{fig:Nb_fits_lt100} show our inferences for the number of \textit{detectable} binary systems (i.e. pairs whose members are separated by more than the adopted resolution limit and both individually brighter than the adopted magnitude limit) within the member sub-population, for dSphs where the input number is larger than 500 and smaller than 200, respectively.  While not explicitly a free parameter, the number of detectable binaries is a straightforward combination of the inferred number of member stars (brighter than the adopted magnitude limit), the inferred member binary fraction, and the integral of the inferred member binary separation function over separations larger than the adopted resolution limit.  

For the samples having more than 500 detectable binary systems, Figures \ref{fig:g1_fits} and \ref{fig:g2_fits} display inferences for the `inner' and `outer' power-law indices, $\gamma_1$ and $\gamma_2'\equiv -\gamma_2/(\gamma_2-2)$, respectively.  Figure \ref{fig:sb_fits} displays results for the break separation, $s_b$, and Figure \ref{fig:L_fits} shows results for the smoothing parameter, $\Lambda$.  

We find that our posteriors generally track the input values.  For the number of detectable binary systems, this success holds even as the number of detectable binaries approaches zero (Figure \ref{fig:Nb_fits_lt100}), providing some reassurance that we are not prone to making spurious detections of binary systems.  Of course, lack of detectable binaries precludes meaningful inference of the binary separation function; indeed, as the number of detectable binaries decreases toward zero, our posteriors for separation function parameters become dominated by the priors listed in Table \ref{table:fitparams}. 

\begin{figure*}
  \centering
  \includegraphics[width=\textwidth]{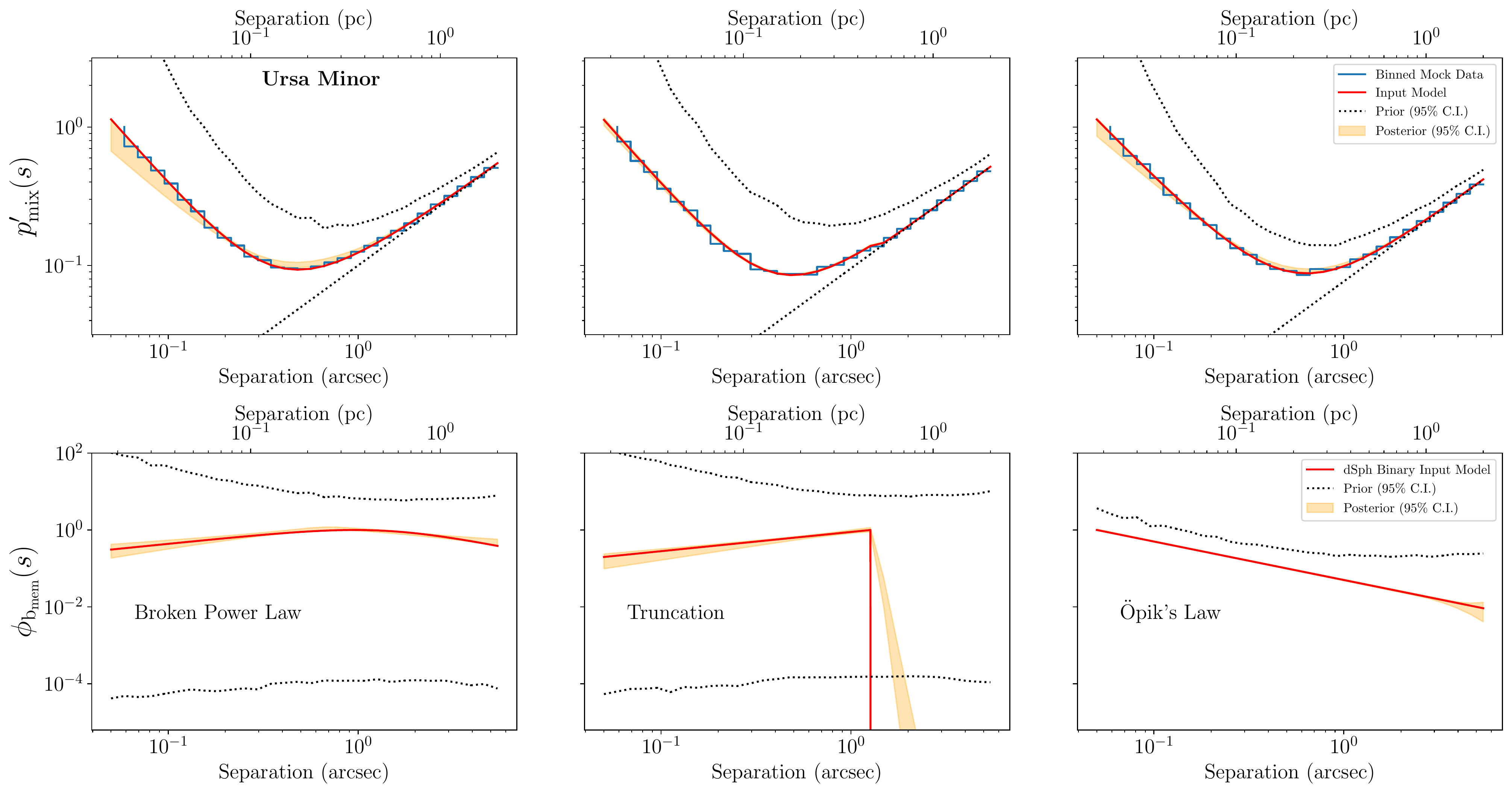}  
\caption{\scriptsize Marginal density of pair separations, from mock data sets drawn for Ursa Minor, with input binary separation function for dSph member stars following a truncated power law (left), broken power law (center), and {\"O}pik's law (right).  In the top row, histograms depict all pair separations regardless of physical association.  Overplotted are the input model (red) and 95\% credibility intervals from the prior probability distribution (enclosed by dashed black lines) and the posterior probability distribution (orange band) that we obtain by fitting the \textit{joint} distribution of pair separations and object positions.  Distributions in bottom panels pertain only to separations between physically-associated pairs within the dSph member sub-population.}
\label{fig:CI_Plots_UMi}
\end{figure*}

\begin{figure*}
  \centering
  \includegraphics[width=\textwidth]{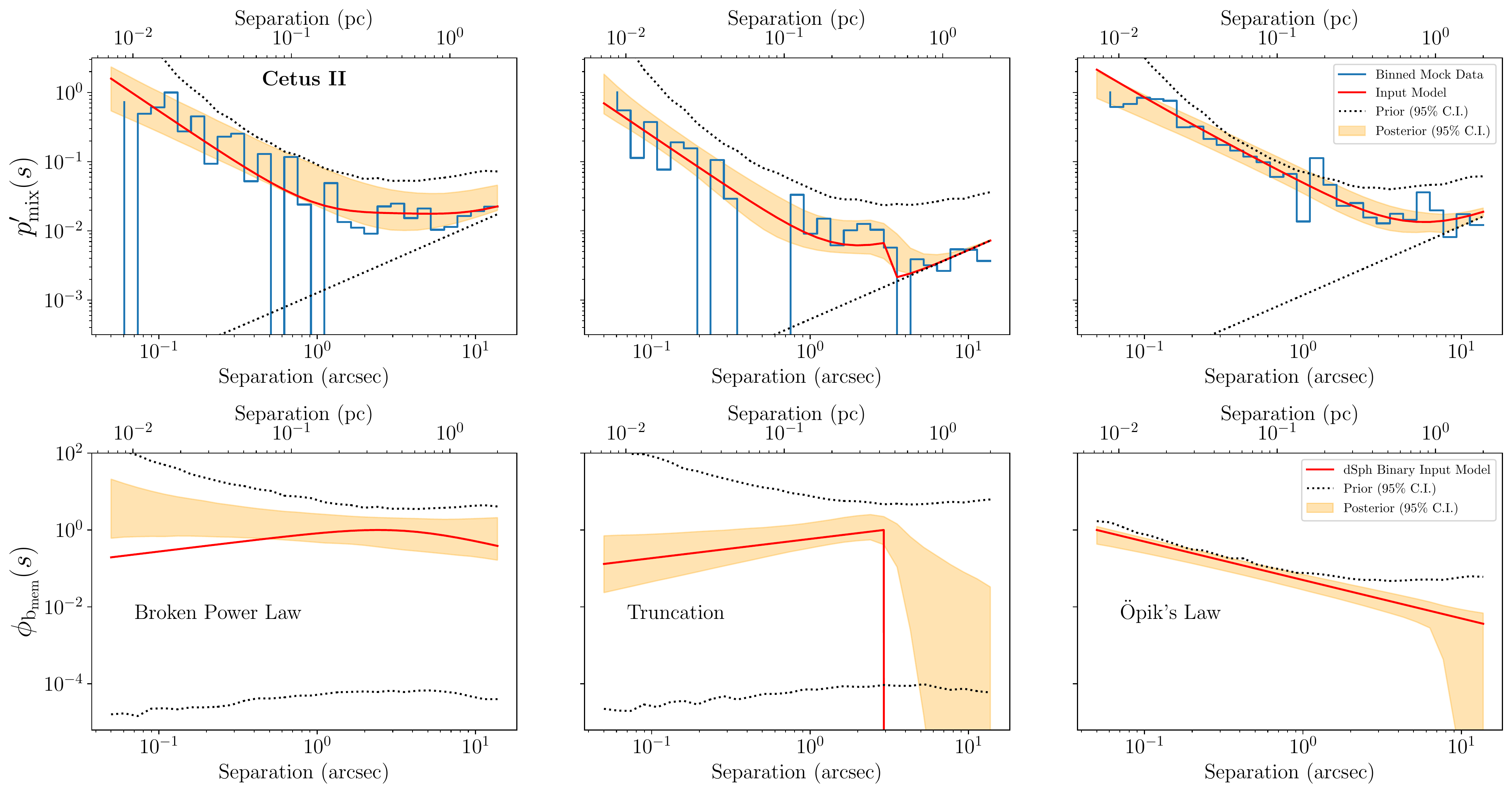}  
\caption{\scriptsize  Same as Figure \ref{fig:CI_Plots_UMi}, but for mock data sets for Cetus II.}
\label{fig:CI_Plots_CetusII}
\end{figure*}

Figures \ref{fig:CI_Plots_UMi} \& \ref{fig:CI_Plots_CetusII} show how our fitted models compare directly to the mock separation data generated for Ursa Minor and Cetus II, chosen to represent cases that are rich and poor, respectively, in detectable binary systems.  Left, center and right panels show marginal densities of pair separations, with input separation functions for member stars following the broken power law (left), truncated power law (center) and {\"O}pik's law (right).  Overplotted are the input model (red curve), as well as 95\% intervals depicting the prior (dotted black curves) and posterior (orange interval) probability distributions.  Top panels show the total separation function observed for all pairs, $p'_{\rm mix}(s)$, regardless of whether the objects are physically associated.  Bottom panels depict only the separations between physically-associated binaries within the member sub-population, $\phi_{\rm mem}(s)$.   

We see that for the mock Ursa Minor, which contains $\ga 3000$ detectable binaries, the posterior intervals for $p'_{\rm mix}(s)$ are difficult to distinguish by eye from the input models.  For the mock Cetus II, which contains $\sim 10-100$ detectable binaries depending on the input model, the posterior intervals for $p'_{\rm mix}(s)$ are obviously wider than for UMi, but still in good agreement with the input models, and indicative that even data sets containing $\mathcal{O}(10)$ detectable binaries can be informative about the separation function.

We now discuss behaviors that are specific to each of the three input models for the member binary separation function.  
\subsection{Broken Power Law}
\begin{figure}
  \centering  
  \includegraphics[width=0.5\textwidth]{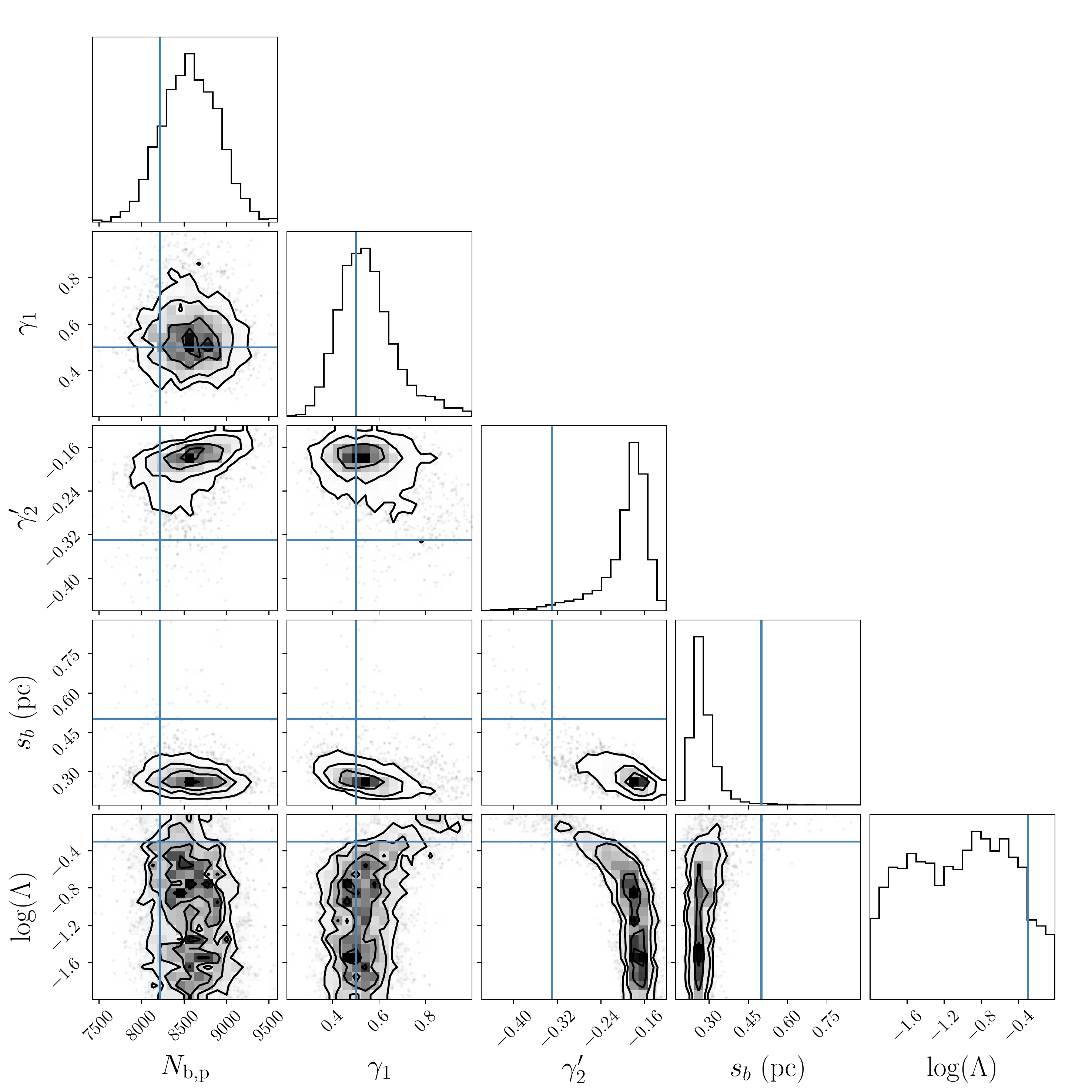}
  \caption{\scriptsize Example of multi-dimensional posterior probability distributions inferred for parameters that specify the binary separation function for the dSph member sub-population, from the mock UMi data set with broken power law input model.  Contours represent 68\%, 95\%, 99\% credible intervals, blue crosshairs marking input values.}
  \label{fig:corner_bpl}
\end{figure}

When the input binary separation function follows the (untruncated) broken power law, we find that the posteriors for both the inner and outer power law indices are generally consistent with input values of $\gamma_1=+0.5$ and $\gamma_2=-1$ (triangular markers in Figures \ref{fig:Nb_fits} - \ref{fig:L_fits}).  Moreover, when the number of detectable binaries exceeds a few hundred, there is sufficient precision to detect the transition around the break separation, as can be seen in the simultaneous constraints on $\gamma_1$, $\gamma_2$ and $s_b$.  However, in modeling the transition there is some degeneracy amongst the outer power-law index, the break separation and the smoothing parameter.  This degeneracy can be seen in Figure \ref{fig:corner_bpl}, which displays, as a representative example, the multi-dimensional posterior that we obtain for the mock data set for Ursa Minor, in the case of input separation function given by the broken power law.  

\begin{figure}
  \centering  
  \includegraphics[width=0.5\textwidth]{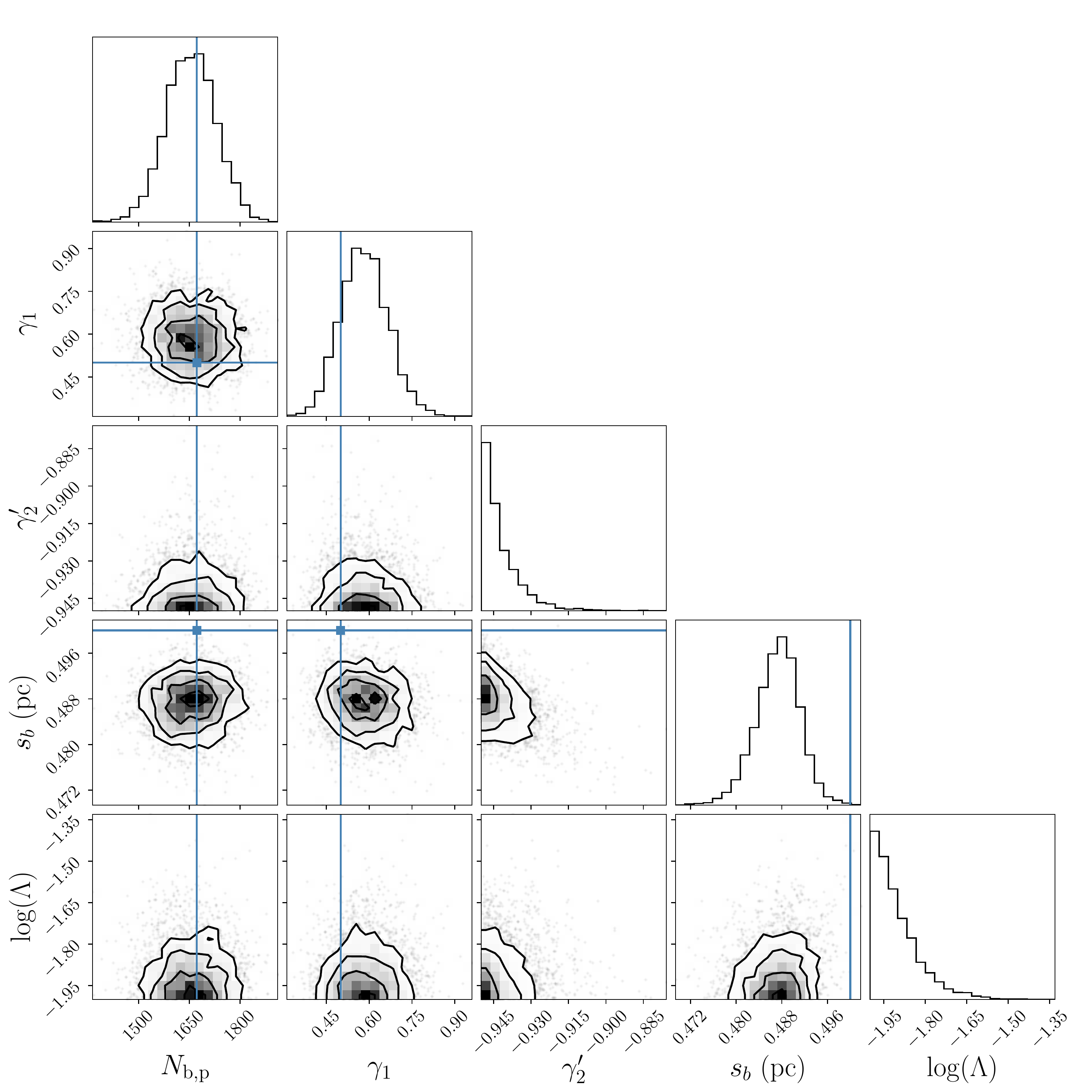}
  \caption{\scriptsize Same as Figure \ref{fig:corner_bpl}, but for the truncated power law as the input model.  }
  \label{fig:corner_trunc}
\end{figure}

\begin{figure}
  \centering  
  \includegraphics[width=0.5\textwidth]{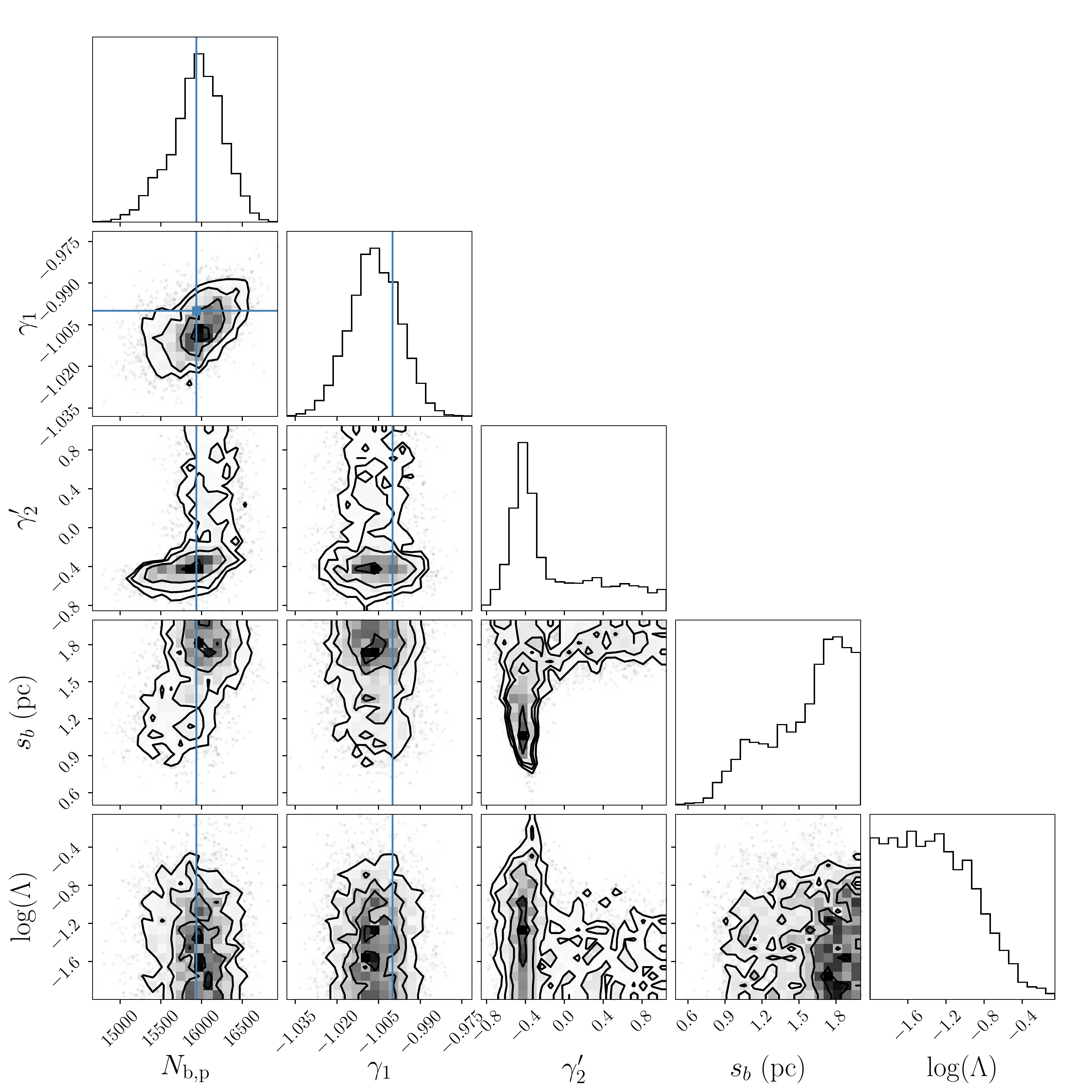}
  \caption{\scriptsize Same as Figure \ref{fig:corner_bpl}, but for {\"O}pik's law as the input model.}
  \label{fig:corner_opik}
\end{figure}

\subsection{Truncated power law}
When the input model follows the truncated power law, we see that our posteriors can accurately recover the abrupt transition from $\gamma_1=+0.5$ to $\gamma_2=-\infty$ (circular markers in Figures \ref{fig:Nb_fits} - \ref{fig:L_fits}).  Here we find that the inner power-law index tends to be tightly constrained around the input value of $\gamma_1=+0.5$.  Sensitivity to a sharp truncation is reflected in the posteriors for the outer power-law index, the smoothing parameter and the break separation.  The posterior for $\gamma_2'$ pushes up against the edge of the prior at $\gamma_2'=-1$, corresponding to $\gamma_2=-\infty$ (since the input value is at the edge of the prior, no finite posterior interval will include the input value).  Likewise, the posterior for the smoothing parameter pushes against the lower edge of its prior, indicating a sharp transition.  As a result, the break separation is tightly constrained around its input value.  Again to provide a  representative example, Figure \ref{fig:corner_trunc} shows an the multi-dimensional posterior obtained for the mock Ursa Minor, with the input separation function given by the truncated power law.  

\subsection{{\"O}pik's Law}
Finally, when the input model follows {\"O}pik's law, our posteriors accurately recover the unbroken power-law index of $\gamma_1=-1$, albeit via a combination of the now-redundant index parameters $\gamma_1$ and $\gamma_2$ (square markers in Figures \ref{fig:Nb_fits} - \ref{fig:L_fits}).  Figure \ref{fig:corner_opik} again shows the multi-dimensional posterior obtained for Ursa Minor, only with the input separation function following {\"O}pik's law.  The index parameters $\gamma_1$ and $\gamma_2'$ both have prominent peaks at values consistent with {\"O}pik's law ($\gamma_1=-1$, $\gamma'_2=-1/3$); however, depending on whether the break separation parameter is smaller than $s_{\rm min}$, larger than the maximum separation $s_{\rm max}$, or somewhere in between, the separation function between $s_{\rm min}$ and $s_{\rm max}$ is controlled, respectively, by $\gamma_2$, $\gamma_1$, or a combination wherein both have values corresponding to {\"O}pik's law.  This behavior is reflected in the fact $\gamma_2$ is tightly constrained around a value of $-1$ when the break separation is small, but virtually unconstrained when the break separation is large.

\section{Summary and Discussion}\label{sec:Conclusion}

We have introduced and developed new analytical tools for the detection and characterization of wide binary systems from catalogs of 2D object positions.  Specifically, we have derived general formulae for calculating the conditional separation function $\mu(s|\vecr)$, the joint position-separation function $\psi(s,\vecr)$, and the marginal separation function $\phi(s)$, for any population of objects with specified position function $\Sigma(\vecr)$.   Moreover, the  object population can be a mixture of an arbitrary number of sub-populations, each of which can have some position-dependent fraction of its objects split into binary systems that follow arbitrary and position-dependent internal separation functions $\phi_{\rm b}(s|\vecr)$.  We have derived analytic separation functions  under conditions that surface density follows a circularly-symmetric Plummer profile, and/or a uniform distribution within a finite circular field.  We have used these results to analyze mock stellar-position data sets that we generated to mimic the observed structural parameters of the Milky Way's known dSph satellites, inserting wide binary populations by hand according to different separation functions that represent different scenarios for wide binary evolution.  We have demonstrated the ability to recover input parameters that govern  binary separation functions, even when the observed sample contains as few as $\mathcal{O}(10)$ binary objects in the member sub-population.  

All of this bodes well for the study of wide binaries within dwarf galaxies, which should become feasible with the launch of next-generation space missions (e.g., James Webb Space Telescope, Nancy Grace Roman Space Telescope) that can provide photometric depth and angular resolution similar to those we have adopted when constructing our mock data sets.  

On the other hand, while the binary separation function for foreground nonmembers within our mock data sets are motivated by recent observational results \citep{tian2019separation}, at present we have no way of knowing whether the binary fractions and separation functions adopted for the mock dSph member populations are realistic, optimistic, or even pessimistic regarding the number of detectable binary systems.  Given this basic uncertainty, the properties of our mock dSph binary populations should not be interpreted as forecasts.  Rather, our adoption of fixed binary fractions are intended, given the range of luminosities and distances within the population of Milky Way dSphs, merely to provide mock data sets containing a wide range of numbers of detectable binaries for testing our methodology.   Reassuringly, we have found that when the number of detectable binaries approaches zero, our modeling accurately recovers that input; thus in the case that real dSphs contain few or no wide binaries that are detectable in future observations, we expect to be able to place limits on the underlying populations.  Such limits can then be useful for constraining wide binary formation/evolution mechanisms and/or dark matter models.    

Some additional caveats apply to the results from our analysis of mock data sets.  While the mock data sets are intended to be realistic in terms of structural parameters and magnitude and resolution limits, they do have some idealized features that will not hold in real observational data sets.  First, the mock dSph surface density functions are all  circularly symmetric, whereas observed dSphs typically have ellipticities $0.1\la \epsilon\la 0.5$ \citep[][and references therein]{mcconnachie12}.  In principle, our analysis need not assume circular symmetry, as the separation functions in Equations  \ref{eq:mu_general} - \ref{eq:phi} hold for arbitrary surface density functions.  In practice, relaxing the modeling assumption of circular symmetry may incur significant additional computational expense, as the joint density $p(s,\vecr)$ may no longer be expressed analytically.

Second, when generating the mock data sets, we have assumed that  resolution and magnitude limits are independent.  In reality, the minimum separation between two discernible point sources will depend on the magnitudes of both objects  \citep[e.g.,][]{chauvin04}.  In order to account for this dependence in real data, provided that it can be quantified using, e.g., artificial star tests, we can update our likelihood function (Equation \ref{eq:likelihood}) to include this dependence directly \citep[see, e.g.,  ][]{el-badry18,tian2019separation}.  We reserve this task for future work, where we plan to test our methodology on more realistic observational catalogs derived  from mock images.  

Third, we reiterate that the approximate likelihood in Equation \ref{eq:likelihood} neglects observational errors associated with measurements of stellar position, as well as covariance amongst the non-independent discrete data points in the data vector.  Comparison of the resulting posteriors to the true input values gives some reassurance that the errors introduced by this approximation are not dramatic---at least not for the input models and mock data sets considered here; however, a more rigorous calculation of the likelihood will necessarily include covariance.  We leave this task for future work that focuses on modeling data from simulated images.

Finally, we note that the formalism presented in Section \ref{sec:method} has potential applications well beyond the analysis of binary star systems.  For example, given a population of surface density $\Sigma(\vecr)$, Equation \ref{eq:phi} can be used to compute directly the probability of random pairings at a given separation, precluding the generation of `random' catalogs when estimating two-point correlation functions.  Furthermore, one could then compare the empirical separation function to the `random' one predicted by Equation \ref{eq:phi} in order to detect in separation space any un-modeled components of the surface density field---e.g., localized substructure and/or under-dense regions.

\acknowledgments
We thank Andrew Pace for providing valuable advice and input.  We thank Lachlan Lancaster, Eric Bell, Anil Seth and Benjamin Williams for providing helpful feedback.  M.G.W. acknowledges support from National Science Foundation grants AST-1813881 and AST-1909584, and by a grant from the McWilliams Center for Cosmology at Carnegie Mellon University and the Pittsburgh Supercomputing Center.  Support for this work was provided in part by the WFIRST Infrared
Nearby Galaxies Survey collaboration through NASA contract NNG16PJ28C.

\bibliography{APJbib}{}
\bibliographystyle{aasjournal}

\appendix 
\section{Derivation of Joint Position-Separation Function}
\label{app:1}

The joint position-separation function for countable items, $\psi'(s,\vecr)$, can be written as the sum of contributions from all combinations of pairings between the different categories of countable items:
\beq
    \psi'(s,\vecr)=\sum_{m}\sum_n\psi'_{m,n}(s,\vecr),
\eeq
where categorical variables $m,n$ both  take values of `singles', `primaries' from binary objects, and `secondaries' from binary objects.  Contributions from pairings between primaries and secondaries depend on whether or not both paired items belong to the same binary object.  All pairings between items \textit{not} belonging to the same binary object contribute as (see Equation \ref{eq:psi})
\begin{equation}
    \psi'_{m,n}(s,\vecr)=\Sigma'_m(\vecr)\,\mu'_n(s|\vecr);\,\,\,m,n \mathrm{\,from\, different\, objects},
    \label{eq:psi_different}
\end{equation}
where $\Sigma_m'(\vecr)$ is the surface number density of countable items in category $m$,  and $\mu'_n(s|\vecr)\equiv \int_0^{2\pi}\Sigma_n'(\vecr+\Delta\vecr)\,s\,\deriv\alpha$ is the conditional separation function for countable items from category $n$.  Pairings between primaries and secondaries from the same binary object contribute as 
\begin{equation}
    \psi'_{m,n}(s,\vecr)=\Sigma'_m(\vecr)\,\phi_{\rm b}(s|\vecr);\,\,\,m,n \mathrm{\,from\,same\,object},
    \label{eq:psi_same}
\end{equation}
where $\phi_{\rm b}(s|\vecr)$ is the internal separation function for the binary pairs, normalized to $\int\phi_{\rm b}(s|\vecr)\,\deriv s=1$.

The surface number density of `single' countable items is  $\Sigma_{\rm singles}'(\vecr)=\bigl (1-f_{\rm b}(\vecr)\bigr )\,\Sigma(\vecr)$, where $\Sigma(\vecr)$ is the surface number density of statistically-independent objects (a binary system is one object containing two countable items).   The surface number densities of both `primary' and `secondary' countable items are approximately\footnote{This approximation holds as long as binary separations are small compared to the scale over which $f_{\rm b}(\vecr)\,\Sigma(\vecr)$ changes. } $\Sigma'_{\rm pri}(\vecr)\approx\Sigma'_{\rm sec}(\vecr)\approx f_{\rm b}(\vecr)\,\Sigma(\vecr)$.  Likewise, the conditional separation functions of singles, primaries and secondaries are  $\mu'_{\rm singles}(s|\vecr)=\bigl (1-f_{\rm b}(\vecr)\bigr )\,\mu(s|\vecr)$ and $\mu'_{\rm pri}(s|\vecr)\approx \mu'_{\rm sec}(s|\vecr)\approx f_{\rm b}(\vecr)\,\mu(s|\vecr)$.  

Table \ref{tab:app} lists the  contribution to $\psi'(s,\vecr)$ from each possible combination of categorical pairs.  The sum is (Equation \ref{eq:psiprime})  
\beq
    \psi'(s,\vecr)\approx\bigl (1+f_{\rm b}(\vecr)\bigr )^2\,\psi(s,\vecr)+2f_{\rm b}(\vecr)\,\,\Sigma(\vecr)\,\phi_{\rm b}(s|\vecr).
\eeq

\begin{table}[b]
    \centering
    \caption{Contributions to $\psi'(s,\vecr)$ from all possible pair combinations of item categories.}
    \begin{tabular}{llll} 
      \hline \hline 
      $m$ & $n$ & $\psi'_{m,n}(s,\vecr)$\\ 
      \hline
      single&single&$\bigl (1-f_{\rm b}(\vecr)\bigr )^2\,\psi(s,\vecr)$\\
      single&primary&$\approx \bigl (1-f_{\rm b}(\vecr)\bigr )\,f_{\rm b}(\vecr)\,\psi(s,\vecr)$\\
      single&secondary&$\approx \bigl (1-f_{\rm b}(\vecr)\bigr )\,f_{\rm b}(\vecr)\,\psi(s,\vecr)$\\
      primary&single&$\approx\bigl (1-f_{\rm b}(\vecr)\bigr )\,f_{\rm b}(\vecr)\,\psi(s,\vecr)$\\
      primary&primary&$\approx\bigl (f_{\rm b}(\vecr)\bigr )^2\,\psi(s,\vecr)$\\
      primary&secondary---same object&$\approx f_{\rm b}(\vecr)\,\Sigma(\vecr)\,\phi_{\rm b}(s|\vecr)$\\
      primary&secondary---different object&$\approx\bigl (f_{\rm b}(\vecr)\bigr )^2\,\psi(s,\vecr)$\\
      secondary&single&$\approx\bigl (1-f_{\rm b}(\vecr)\bigr )\,f_{\rm b}(\vecr)\,\psi(s,\vecr)$\\
      secondary&primary---same object&$\approx f_{\rm b}(\vecr)\,\Sigma(\vecr)\,\phi_{\rm b}(s|\vecr)$\\
      secondary&primary---different object&$\approx \bigl (f_{\rm b}(\vecr)\bigr )^2\,\psi(s,\vecr)$\\
      secondary&secondary&$\approx\bigl (f_{\rm b}(\vecr)\bigr )^2\,\psi(s,\vecr)$\\
      \end{tabular}
    \label{tab:app}
\end{table}

\vspace{5mm}
\end{document}